\documentclass[journal]{IEEEtran}
\usepackage{bm}
\usepackage{amssymb}
\usepackage{amsfonts}
\usepackage{amsmath}
\usepackage{algorithm}
\usepackage{balance}
\usepackage{graphicx,subfig,cite,multicol,multirow,diagbox,booktabs,array}
\usepackage{epstopdf}

\usepackage{color}
\usepackage{float}
\usepackage{stfloats}
\usepackage{makecell}
\ifCLASSINFOpdf
\else
\fi
\begin{document}
\title{DOA Estimation for Hybrid Massive MIMO Systems using Mixed-ADCs: Performance Loss and Energy Efficiency}

\author{Baihua Shi, Qi Zhang, Rongen Dong, Qijuan Jie, Shihao Yan, Feng Shu, and Jiangzhou Wang,~\IEEEmembership{Fellow,~IEEE}
\thanks{This work was supported in part by the National Natural Science Foundation of China (Nos.U22A2002, 61972093 and 62071234), Hainan Province Science and Technology Special Fund (ZDKJ2021022), and the Scientific Research Fund Project of Hainan University under Grant KYQD(ZR)-21008.}
\thanks{B. Shi is with the School of Electronic and Optical Engineering, Nanjing University of Science and Technology, Nanjing 210094, China.}
\thanks{F. Shu is with the School of Information and Communication Engineering, Hainan University, Haikou, 570228, China and also with the School of Electronic and Optical Engineering, Nanjing University of Science and Technology, Nanjing, 210094, China. (e-mail: shufeng0101@163.com)}
\thanks{Q. Zhang, R. Dong and Q. Jie are with the School of Information and Communication Engineering, Hainan University, Haikou 570228, China.}
\thanks{S. Yan is with the School of Science and Security Research Institute, Edith Cowan University, Perth, WA 6027, Australia.}
\thanks{J. Wang is with the School of Engineering and Digital Arts, University of Kent, Canterbury CT2 7NT, U.K. (e-mail: j.z.wang@kent.ac.uk).}
}
\maketitle

\begin{abstract}
	Due to the power consumption and high circuit cost in antenna arrays, the practical application of massive multiple-input multiple-output (MIMO) in the sixth generation (6G) and future wireless networks is still challenging. Employing low-resolution analog-to-digital converters (ADCs) and hybrid analog and digital (HAD) structure is two low-cost choice with acceptable performance loss.
	In this paper, the combination of the mixed-ADC architecture and HAD structure employed at receiver is proposed for direction of arrival (DOA) estimation, which will be applied to the beamforming tracking and alignment in 6G. By adopting the additive quantization noise model, the exact closed-form expression of the Cram\'{e}r-Rao lower bound (CRLB) for the HAD architecture with mixed-ADCs is derived. Moreover, the closed-form expression of the performance loss factor is derived as a benchmark. In addition, to take power consumption into account, energy efficiency is also investigated in our paper. The numerical results reveal that the HAD structure with mixed-ADCs can significantly reduce the power consumption and hardware cost. Furthermore, that architecture is able to achieve a better trade-off between the performance loss and the power consumption. Finally, adopting 2-4 bits of resolution may be a good choice in practical massive MIMO systems.
\end{abstract}

\begin{IEEEkeywords}
	DOA estimation, hybrid analog and digital, mixed-ADC, CRLB, energy efficiency
\end{IEEEkeywords}

\IEEEpeerreviewmaketitle

\section{Introduction}
As an important technique in wireless communication, direction of arrival (DOA) estimation has attracted widely attention \cite{DOA,chenDOA2022wcl,zhuang2020machine}. DOA estimation has lots of applications, like arrival of angle (AOA) location, sonar, rescue, and tracking of objects. In modern and future communication, DOA estimation may have more potential applications: unmanned aerial vehicle (UAV) communications \cite{zengUAV}, directional modulation (DM) systems \cite{shuDMRIS2021tcom}, intelligent reflecting surface (IRS) communications \cite{HongIRS,WuIRS}, internet of things (IoT) \cite{KaurIOT}, and millimeter-wave-based massive multiple input multiple output (MIMO) and so on in the sixth generation (6G) \cite{dongRIS2022ojcs,jieDOA2022wcl}.

It is known that the most methods of DOA estimation are based on spatial spectrum. The multiple signal classification (MUSIC) algorithm was proposed by Schmidt in \cite{MUSIC}. However, the MUSIC method needs to obtain a spectrum of all searching angles, which is of high-complexity. Thus, to refrain from the spectral search in the MUSIC method algorithm, root-MUSIC was proposed in \cite{rootmusic}, which replaces the spectral search with solving polynomial zeros. In \cite{ESPRIT}, authors  proposed another famous search-free method, estimation of signal parameters via rotational invariance technique (ESPRIT).  Afterwards, many root-MUSIC-based and ESPRIT-based algorithm were proposed. To improve the performance of the ESPRIT , a total least-squares approach, TLS-ESPRIT, was proposed in \cite{TLS_ESPRIT}. 
In order to reduce the complexity of the Root-MUSIC, unitary root-MUSIC (U-root-MUSIC) and real-valued root-MUSIC (RV-root-MUSIC) were proposed in \cite{U_rootmusic} and \cite{RV_rootmusic}, respectively. Recently, more and more concern has been attracted to the DOA estimation for massive MIMO systems. In \cite{chengDOA}, the DOA estimation in two-dimensional (2-D) massive MIMO systems was investigated with no access to the number of signals, path gain correlations etc. Although the performance of spatial-spectrum-based methods is very satisfactory, the complexity is still high, especially in massive MIMO systems. 

The practical implementation of massive MIMO systems is still difficult. Each antenna is connected to a radio frequency (RF) chain in massive MIMO systems. And, there is an ADC and a digital-to-analog converter (DAC) in every RF chain. It is well-known that the hardware cost and power consumption of ADCs and DACs will increase linearly with the bandwidth and increase exponentially with the number of the quantization bits \cite{LeeADC}. Thus, as the number of antennas increases greatly in massive MIMO systems, the power consumption and hardware cost increase rapidly.

To solve this problem, Adopting hybrid analog and digital (HAD) structure is a promising solution. In this structure, Multiple antennas are connected to one RF chain, resulting in the decrease of RF chain number in the antenna array. In recent years, many researchers have focused on the DOA estimation in HAD structure \cite{shiHADDOA2022scis,shuDOA,wuDOA,fanDOA,LiDOA,huDLDOA,zhuang2020machine}. In \cite{shuDOA}, four low-complexity and high-resolution algorithm were proposed. And, the exact closed-form expression of the CRLB for HAD structure was derived. To tack the phase-ambiguity in partially-connected HAD structure, authors redesigned the analog phase shifts and estimated the DOA through two steps in \cite{wuDOA}. In \cite{fanDOA}, a 2-D discrete Fourier transform (DFT) based algorithm was proposed. Moreover, authors derived the corresponding CRLB of the channel gain and joint DOA estimation. In \cite{LiDOA}, authors proposed a beam sweeping algorithm, reconstructing the spatial covariance matrix of the HAD architecture. Then, in \cite{huDLDOA}, a deep-learning-based DOA estimation method for the HAD architecture with a uniform circular array (UCA) was proposed. However, DOA estimation for the HAD architecture with the UCA is still an open problem.

Another promising solution is to substitute the low-resolution ADCs for the high-resolution ADCs. However, the signals quantified by low-resolution ADCs are nonlinear due to the low-precision  quantization, which is hard to analyse.
In \cite{SinghLowADC}, authors presented that using the additive quantization noise model (AQNM) is able to dispel the low-resolution ADCs' distortion. Then, many researches have been focused on the low-resolution structure. Furthermore, authors revealed that, resulting from adopting low-resolution ADCs, the achievable rate will decrease but that can made up by increasing the number of antennas.
Authors investigated the uplink achievable rate for massive MIMO systems when finite resolution ADCs and the common maximal-ratio combining technique were employed at the receive array in \cite{FanLowADC}.
A multipair massive MIMO two-way relay network was considered in \cite{JinLowADC}, where a relay station with a large number of antennas serves multiple pairs of users.
In \cite{KongLowADC}, a multipair full-duplex massive MIMO relaying system with low-resolution ADCs at both the relay and receivers was considered. Moreover, authors derived the achievable rates for that case with a finite number of antennas at users.
In \cite{XuLOWADC}, authors proposed an algorithm for the physical layer security in the massive MIMO system, which was equipped with the analog phase shifters, finite-quantized digital phase shifters and low-resolution ADCs.
Authors proposed a mixed one-bit array in \cite{wangDOA2018twc}. Furthermore, compressive sensing based methods were proposed for DOA estimation.
Then, the performance loss of DOA estimation with low-resolution ADCs was firstly investigated in \cite{Shiarxiv}. In addition, authors discussed the determination of quantization bits. However, a novel DOA estimation method of the low-resolution architecture has not been proposed.

But the low-resolution structure still has some challenges, including time-frequency synchronization, achievable rate and so on \cite{LiangMIXED,Zhangtcom}.
In order to solve the above problems, Mixed-ADC architecture, as a new architecture, was proposed to replace the low-resolution structure in \cite{LiangMIXED}. In that structure, some RF chains connect the high-resolution ADCs and the others are connected with the low-resolution ADCs. Thus, more and more scholars have fascinated by the mixed-ADC architecture.
In \cite{accessMIXED}, authors investigated the spectral efficiency of massive MIMO systems with mixed-ADCs and proved that the mixed-ADC architecture has a considerable spectral efficiency with a much lower power consumption and circuit cost.
In \cite{Zhangtcom}, authors derived the closed-form expressions of the achievable rate for  multipair massive MIMO relaying systems with mixed-ADCs/DACs. What's more, an efficient power allocation algorithm was proposed.
In \cite{shiDOAmixedADC}, the performance loss of the DOA estimation for the massive MIMO receive array with mixed-ADCs was investigated. And, authors showed that the mixed-ADC architecture can achieve a good trade-off between the performance loss and power consumption.

In this paper, we consider the DOA estimation in the hybrid analog and digital massive MIMO system with mixe-ADCs. To the best of our knowledge, that case has not been investigated. This study aims  to analyse the performance loss and energy efficiency of the DOA estimation for the HAD structure with mixed-ADCs. We show that this new structure can approach the performance of the ideal unquantized system with much lower power consumption.
The main contributions of this paper are summarized as follows:
\begin{enumerate}
	\item Considering the nonlinear error caused by the low-resolution ADCs, we build up the system model of the DOA estimation for the HAD architecture with mixed-ADCs by resorting to the AQNM. For the comparison with the conventional full digital structure with pure high-resolution ADCs, we employ the design of analog phase shifters in \cite{shuDOA}. Based on that, the root-MUSIC-based method in the HAD structure is proved that it can be used in the HAD architecture with mixed-ADCs without any other modification in the numerical simulation.
	\item To eliminate ambiguity in single time block for HAD structure, we redesign the analog beamforming matrix. The proposed method is able to estimate the direction in time block. Furthermore, proposed method is robust due to all directions are covered by newly designed analog beamforming matrix.
	\item To obtain an accurate performance loss of the DOA estimation for the HAD structure with mixed-ADCs, we define the performance loss factor. Then, we derive the closed-form expression of the CRLB for the HAD structure with mixed-ADCs by the statistic theory and matrix theory. Furthermore, the closed-form expression of the performance loss factor is also derived. Our results show that the HAD structure with mixed-ADCs causes more performance loss than the full digital structure with mixed-ADCs. However, that shortcoming can be made up by increasing the low-resolution ADCs' quantization bits and the number of high-resolution ADCs.
	\item Finally, in order to assess a thorough investigation on the DOA estimation for the HAD structure with mixed-ADCs, it is essential to study the energy efficiency for that architecture. Combined with the performance loss, the simulation results reveal that although the HAD structure with mixed-ADCs has worse performance, its energy efficiency is much higher. It means that the HAD structure with mixed-ADCs is more suitable for the practical massive MIMO systems. In addition, we find that using 2-4 bits of resolutions can strike a better trade-off and adopting ADCs with longer quantization bits just causes little improvement of performance with much higher power consumption.
\end{enumerate}

The rest of this paper is organized as follows. The system model of the DOA estimation for the HAD structure with mixed-ADCs is presented in Section 2. Then, A novel method is proposed in Section 3. The performance loss and energy efficiency are investigated in Section 4. In Section 5, simulation and numerical results are provided to analyze the performance and trade-off. Finally, we come to the conclusion in Section 6.

\emph{Notations:} In this paper, signs $(\cdot)^T$, $(\cdot)^H$, $|\cdot|$ and $\|\cdot\|$ represent transpose, conjugate transpose, modulus and norm, respectively. $\mathbf{x}$ and $\mathbf{X}$ in bold typeface are used to represent vectors and matrices, respectively, while scalars are presented in normal typeface, such as $x$. $\mathbf{I}_M$ represents the $M\times M$ identity matrix, $\mathbf{0}_{a\times b}$ denotes the $a\times b$ matrix of all zeros and $\mathbf{1}_M$ denotes the $M\times 1$ vector of all $1$. Furthermore, $\mathbb{E}[\cdot]$ and $\mathbb{R}[\cdot]$ denotes the expectation operator and the real part  of a complex-valued number, respectively. $\otimes$ denotes the Kronecker product. $\mathbf{diag}(\mathbf{X})$ and $\mathbf{Diag}(\mathbf{x})$ represent a diagonal matrix by keeping only the diagonal elements of matrix $\mathbf{X}$ and a diagonal matrix composed of $\mathbf{x}$.  $\mathbf{Tr}(\cdot)$ and $\lceil\cdot\rceil$ respectively denote the matrix trace and the round up operation. 
\section{System Model}
\begin{figure}
  \centering
  \includegraphics[width=0.48\textwidth]{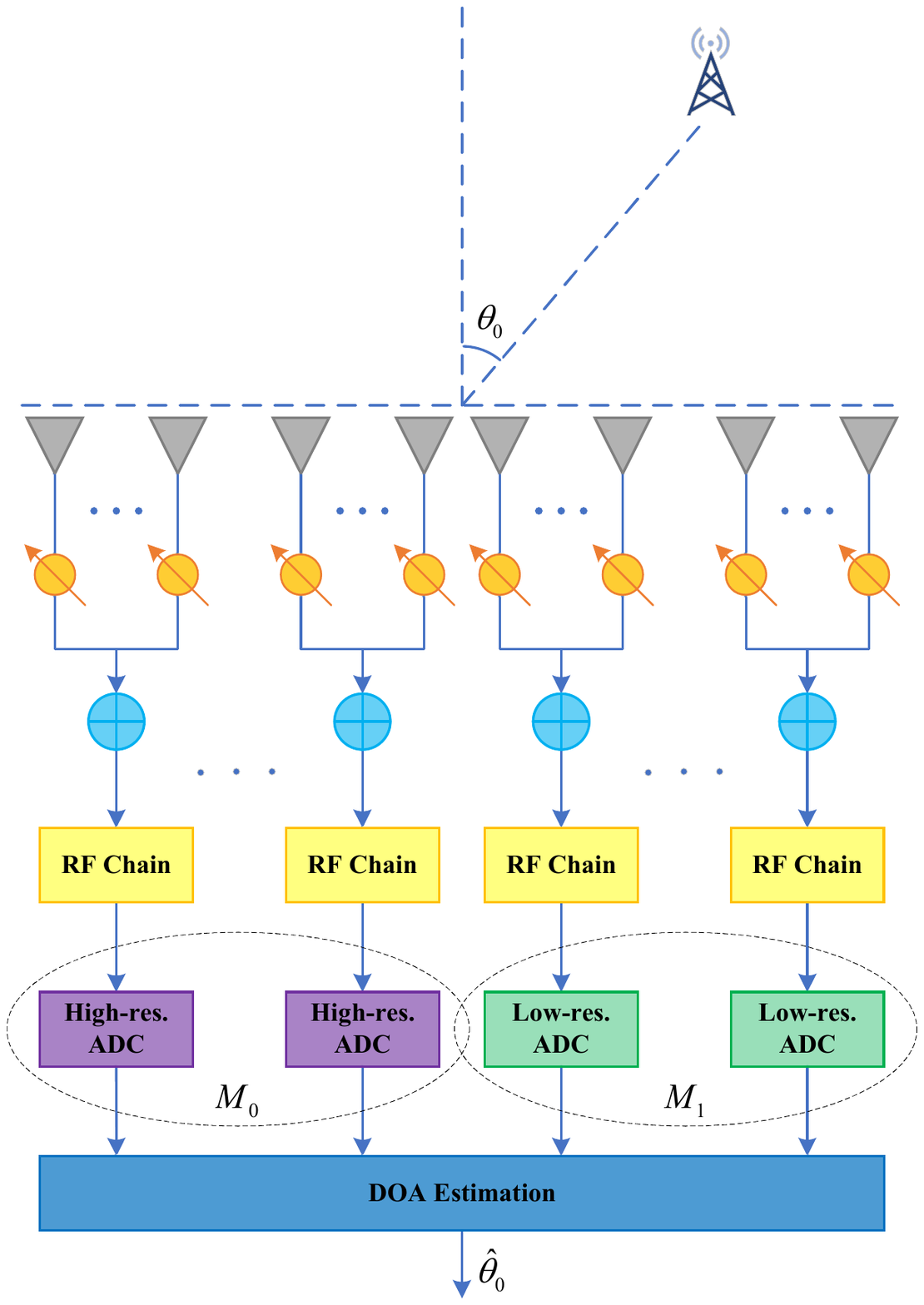}\\
  \caption{System model of the HAD architecture with mixed-ADCs}
  \label{fig_sys_mod}
\end{figure}

As illustrated in Fig.~\ref{fig_sys_mod}, we consider a sub-connected hybrid analog and digital architecture with mixed-ADCs. A narrow-band signal $s(t)e^{j2\pi f_ct}$ is sent from a far-field emitter and strikes the uniformly-spaced linear array (ULA), where $f_c$ is the carrier frequency. The ULA with $M$ antenna elements is divided into $M_s$ subarrsys, and there are $M_a$ antenna elements in the every subarray, $M=M_sM_a$. The propagation delays from the signal to all antenna elements are given by
\begin{table}
\footnotesize
\centering
\caption{Linear Quantization Gain $\alpha$ For Different Quantization Bits ($b<6$)}
\label{tab1}
\scalebox{1.18}{
\begin{tabular}{c|c|c|c|c|c}
\hline
\hline
$b$     & 1     & 2     & 3         & 4         & 5     \\
\hline
$\beta$ & 0.3634& 0.1175& 0.03454   &0.009497   &0.002499     \\
\hline
\hline
\end{tabular}}
\end{table}
\begin{equation}\label{tau}
\tau_m=\tau_0-\frac{d_m}{c}\sin\theta_0,~m=1,2,\cdots,M,
\end{equation}
where $\tau_0$, $d_m$, and $c$ denotes the transmission time from the emitter to the reference point on the array, the distance from the $m$th antenna element to the reference point, and the speed of the light, respectively. $\theta_0$ represents the direction of arrival, which ranges from $-90^\circ$ to $90^\circ$. Thus, the received signal before the analog beamforming (AB) can be expressed as
\begin{equation}\label{x}
\mathbf{x}(t)=\mathbf{a}(\theta_0)s(t)e^{j2\pi f_ct}+\mathbf{w}(t),
\end{equation}
where $\mathbf{w}(t)$ denotes the the additive white Gaussian noise (AWGN). Thus, all elements of $\mathbf{w}(t)$ are independent and identically distributed (i.i.d.) and $\mathbf{w}(t)\sim \mathcal{CN}(0,\mathbf{I}_M)$. $\mathbf{a}(\theta_0)$ is the array manifold, which is given by
\begin{equation}\label{a}
\mathbf{a}(\theta_0)=\left[e^{j  \Psi_{\theta_0} (1)}\ e^{j  \Psi_{\theta_0} (2)}\ \cdots\ e^{j  \Psi_{\theta_0}(M)}\right]^T,
\end{equation}
where
\begin{equation}\label{Psi}
\Psi_{\theta_0}(m) = 2\pi\sin\theta_0\frac{d_{m}}{\lambda},~m=1,2,\cdots,M,
\end{equation}
is the signal phases of $m$th antenna element, corresponding to the propagation delay difference between $m$th antennal element and the reference point. Hence, for the ULA, if we choose the endpoint element as the reference point, $d_m$ is given by
\begin{equation}\label{dm}
d_{m}=(m-1)d,~m=1,2,\cdots,M,
\end{equation}
where $d$ is the antenna spacing. Furthermore, $d_m$ can be written as
\begin{align}\label{dh}
d_{m_s,m_a} = [(&m_s-1)M_a+m_a-1] d, \nonumber\\
&m_s=1,2,\cdots,M_s,~m_a=1,2,\cdots,M_a.
\end{align}
After going through the AB and RF chain, the signal is down-converted from frequency band to baseband, which is given by
\begin{equation}\label{yt}
\mathbf{y}(t)=\mathbf{V}_{A}^H \mathbf{a}(\theta_0)s(t)+\mathbf{w}(t),
\end{equation}
where the AB matrix is defined by
\begin{equation}\label{Va}
\mathbf{V}_{A}=\left[
\begin{array}{cccc}
  \mathbf{v}_{A,1}  & \mathbf{0}        & \cdots & \mathbf{0} \\
  \mathbf{0}        & \mathbf{v}_{A,2}  & \cdots & \mathbf{0} \\
  \vdots            & \vdots            & \ddots & \vdots \\
  \mathbf{0}        & \mathbf{0}        & \cdots & \mathbf{v}_{A,M_s}
\end{array}
\right],
\end{equation}
where
\begin{equation}\label{Vms}
\mathbf{v}_{A,m_s}=\frac{1}{\sqrt{M_a}}\left[ e^{j\omega_{m_s,1}}~e^{j\omega_{m_s,2}}\cdots e^{j\omega_{m_s,M_a}}\right]^T
\end{equation}
is the AB vector of the $m_s$th subarray.

\begin{figure*}[hb] 
	\vspace*{4pt}
	\hrulefill
	\begin{equation}\label{ytt}
		\mathbf{y}(t)=\left[
		\begin{array}{c}
			\mathbf{y}_0(t) \\
			\mathbf{y}_1(t)
		\end{array}
		\right]=\left[
		\begin{array}{cc}
			\mathbf{V}_{A,0}              & \mathbf{0}_{M_0 \times M_1} \\
			\mathbf{0}_{M_1 \times M_0}   & \mathbf{V}_{A,1}
		\end{array}
		\right]
		\left[
		\begin{array}{c}
			\mathbf{a}_0(\theta_0) \\
			\mathbf{a}_1(\theta_0)
		\end{array}
		\right]s(t)+\left[
		\begin{array}{c}
			\mathbf{w}_0(t) \\
			\mathbf{w}_1(t)
		\end{array}
		\right]
	\end{equation}
\end{figure*}
Different form the conventional HAD structure with pure high-resolution ADCs, $M_0$ RF chains are connected with high-resolution ADCs and the other $M_1$ RF chains are connected with low-resolution ADCs in our architecture, which means $M_s=M_0+M_1$. Then, $\kappa\triangleq M_0/M_s~(0\leq\kappa\leq1)$ is defined as the proportion of the high-resolution ADCs. Thus, $\mathbf{y}(t)$ can be divided into two parts as (\ref{ytt}), shown at the bottom of this page.
The received signals passing through high-resolution ADCs can be given by
\begin{align}\label{y0n}
\mathbf{y}_{0}(n)=\mathbf{y}_{0}(t)|_{t=n}=\mathbf{V}_{A,0}^H\mathbf{a}_0 (\theta_0)&s(n)+\mathbf{w}_0(n)\nonumber\\
&n=1,2,\cdots,N
\end{align}
where $N$ is the number of snapshots, and $\mathbf{w}_0(n)\sim \mathcal{CN}(\mathbf{0},\mathbf{I}_{M_0})$ denotes the AWGN vector with i.i.d. components following the distribution $\mathcal{CN}(0,1)$. By resorting to the additive quantization noise model (AQNM) widely adopted in massive MIMO systems \cite{Lowpower,Zhangjsac,jinshilow} , the nonlinear quantization error caused by the low-resolution ADCs can be transmitted into a linear gain with the additional noise.
\begin{align}\label{y1n}
\mathbf{y}_{1}(n)=\mathbb{Q}\{\mathbf{y}_{1}(t)\}=\alpha\mathbf{V}_{A,1}^H\mathbf{a}_1 (\theta_0)&s(n)+\alpha\mathbf{w}_1(n)+\mathbf{w}_q(n)\nonumber\\
&n=1,2,\cdots,N
\end{align}
where $\mathbf{w}_1(n)\sim \mathcal{CN}(0,\mathbf{I}_{M_1})$, $\mathbb{Q}\{\cdot\}$ is the quantization function of the low-resolution ADCs, $\mathbf{w}_q(n)$ is the quantization noise related to the $\mathbf{y}_1(t)$, and $\alpha=1-\beta$ is a linear gain, where $\beta$ denotes the distortion factor of the low-resolution ADCs, defined by
\begin{equation}\label{beta}
\beta=\frac{\mathbb{E}\left[ \|\mathbf{y}-\mathbf{y}_{q}\|^{2}\right]}{\mathbb{E}\left[ \|\mathbf{y}\|^{2}\right]}.
\end{equation}
The exact values of $\beta$ with different quantization bits are listed in Table \ref{tab1}. When $b>5$, the distortion factor $\beta$ can be approximated as
\begin{equation}\label{beta6}
\beta\approx\frac{\sqrt{3}\pi}{2}\cdot 2^{-2b},~b\geq 6.
\end{equation}
Then, the covariance matrix of $\mathbf{w}_q(n)$ can be expressed as
\begin{equation}\label{R_wh}
\mathbf{R}_{\mathbf{w}_q}=\alpha\beta \mathbf{diag}(\sigma_s^2\mathbf{V}_{A,1}^H\mathbf{a}_{1}(\theta) \mathbf{a}_{1}^H(\theta)\mathbf{V}_{A,1}+\mathbf{I}_{M_1}).
\end{equation}
By combining (\ref{y0n}) and (\ref{y1n}),
the total quantized signal is given by
\begin{align}\label{yh}
\mathbf{y}(n)&=\left[
\begin{array}{c}
\mathbf{y}_{0}(n) \\
\mathbf{y}_{1}(n)
\end{array}
\right]\nonumber\\
&\approx
\left[
\begin{array}{c}
\mathbf{V}_{A,0}\mathbf{a}_{0}(\theta_0)\mathbf{s}(n)+\mathbf{w}_{0}(n) \\
\alpha\mathbf{V}_{A,1}\mathbf{a}_{1}(\theta_0)\mathbf{s}(n)+\alpha \mathbf{w}_{1}(n)+\mathbf{w}_{q}(n)
\end{array}
\right].
\end{align}

\section{Proposed Single Time Block Estimator for HAD structure}
\begin{figure}
	\centering
	\includegraphics[width=0.48\textwidth]{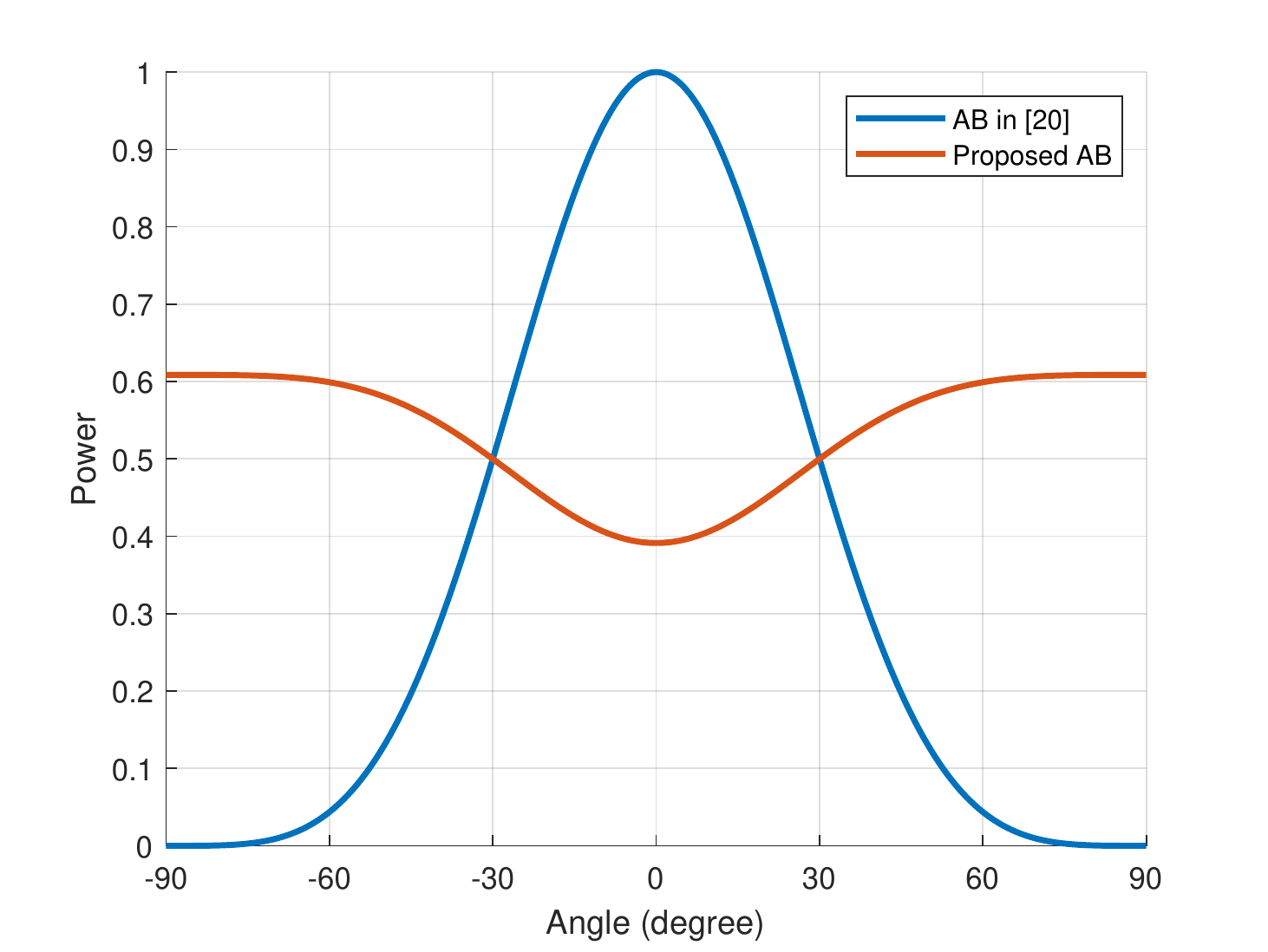}\\
	\caption{Normalized energy of different AB for a HAD structure with $M=16$ and $M_a = 2$}
	\label{fig_ABpower}
\end{figure}
In this section, we design a novel AB to eliminate ambiguity in one time block. Referring to \cite{shiDOAmixedADC}, the subspace-based methods could be applied in mixed-ADC architecture without modification. Thus, for the sake of simplicity, the derivation of proposed method is based on the array only equipped with high-resolution ADCs.

Every subarray is equipped with $M_a$ antennas. Hence, the 3dB beamwidth of subarray can be expressed as
	\begin{equation}\label{theta_3db}
		\theta_{3db} \approx \frac{50.8\lambda}{M_ad}~(^\circ)
	\end{equation}
Note that, in this paper, the range of direction is $[-\pi/2,\pi/2]$. Thus, it is necessary that
$M_s\geq \lceil180/\theta_{3db}\rceil$.
To cover all directions, we design the element of AB by 
	\begin{equation}\label{analog_sub}
		\omega_{m_s,m_a} = -j2\pi \frac{(m_a-1)d\sin\theta_{m_s}}{\lambda}
	\end{equation}
where
	\begin{equation}\label{theta_ms}
		\theta_{m_s} = \frac{m_s\pi}{M_s}-\frac{\pi}{2}-\frac{\pi}{2M_s}.
	\end{equation}
As shown in Fig.~\ref{fig_ABpower} on the next page, compared with the AB in \cite{shuDOA}, the energy of received signals is more even by adopting proposed AB. Since the direction is unknown, proposed AB is more robust.
Then, the received signal of the $m_s$th RF chain is given by
	\begin{align}\label{y_ma}
		y_{m_s}(t) &= \frac{1}{\sqrt{M_a}}\sum_{m_a}^{M_a}e^{-j\frac{2\pi}{\lambda}(m_a-1)d\sin\theta_{m_s}} e^{j\frac{2\pi}{\lambda}(m_a-1)d\sin\theta_0} \nonumber\\ &~~~\times e^{j\frac{2\pi}{\lambda}(m_s-1)M_a d\sin\theta_0}s(t)+w(t) \nonumber\\
		&=\frac{1}{\sqrt{M_a}}e^{j\frac{\pi}{\lambda}(M_a-1)d\sin\theta_0} e^{-j\frac{\pi}{\lambda}(M_a-1)d\sin\theta_{m_s}}\nonumber\\
		&~~~\times\frac{\sin\left(M_a \frac{\pi d}{\lambda}(\sin\theta_{0}-\sin\theta_{m_s})\right)}{\sin\left(\frac{\pi d}{\lambda}(\sin\theta_{0}-\sin\theta_{m_s})\right)} \nonumber\\
		&~~~\times e^{j\frac{2\pi}{\lambda}(m_s-1)M_a d\sin\theta_0} s(t)+w(t)
	\end{align}
Since $\theta_{m_s}$ is known, (\ref{y_ma}) can be written as 
\begin{equation}\label{y_ms}
	y_{m_s}(t) = \delta_{\theta_{m_s}} e^{j\frac{2\pi}{\lambda}(m_s-1)M_a d\sin\theta_0} s(t)+w(t)
\end{equation}
where 
\begin{align}\label{delta}
	\delta_{\theta_{m_s}}=&\frac{1}{\sqrt{M_a}}e^{j\frac{\pi}{\lambda}(M_a-1)d(\sin\theta_0-\sin\theta_{m_s})} \nonumber\\
	&\frac{\sin\left(M_a \frac{\pi d}{\lambda}(\sin\theta_{0}-\sin\theta_{m_s})\right)}{\sin\left(\frac{\pi d}{\lambda}(\sin\theta_{0}-\sin\theta_{m_s})\right)}
\end{align}
is a constant complex number. Thus, the variance of phases in (\ref{y_ms}) is the same as received signals
in conventional ULA. Then, we just need to deal with $delta$. Let us respectively define a digital phase shifter matrix and energy matrix as
	\begin{align}\label{digital_bf}
		\mathbf{V}_{D} = \mathbf{Diag}([e^{j\frac{\pi}{\lambda}(M_a-1)d\sin\theta_{1}}, &e^{j\frac{\pi}{\lambda}(M_a-1)d\sin\theta_{2}},\cdots,\nonumber\\
		&e^{j\frac{\pi}{\lambda}(M_a-1)d\sin\theta_{m_s}}]^T)
	\end{align}
and
	\begin{equation}\label{P_norm}
		\mathbf{P}_{M_s} = \mathbf{Diag}([\sqrt{P_{1}},\sqrt{P_{2}},\cdots,\sqrt{P_{M_s}}]^T)
	\end{equation}
where $P_{m_s} = \mathbb{E}\left[ y_{m_s}(n)^2\right]$.
Hence, when the noise is ignored, the energy-normalization processed signal can be expressed by
	\begin{align}\label{y_final}
		\mathbf{y}_{D}(n) &= \mathbf{P}_{M_s}^{-1}\mathbf{V}_{D}\mathbf{y}(n)=e^{j\frac{\pi}{\lambda}(M_a-1)d\sin\theta_0} \nonumber\\
		&\left[e^{j\frac{2\pi}{\lambda}(1)M_a d\sin\theta_0}, e^{j\frac{2\pi}{\lambda}(2)M_a d\sin\theta_0},\cdots,\right.\nonumber\\
		&~~~~~~~~~~~\left.e^{j\frac{2\pi}{\lambda}(m_s-1)M_a d\sin\theta_0}\right]^T
	\end{align}
It is obvious that (\ref{y_final}) can be solved by subspace-based methods, like root-MUSIC and ESPRIT. However, the ambiguity is also born. By adopting root-MUSIC, We could achieve $M_a$ candidate solutions $\bm{\Theta_c} = [\hat{\theta}_{c,1},\hat{\theta}_{c,2},\cdots,\hat{\theta}_{c,M_a}]^T$ as follows
	\begin{equation}\label{theta_candidate}
		\hat{\theta}_{c,i}=\arcsin \left( \frac{\lambda\arg \hat{z}}{2\pi M_a d} + \frac{\lambda i}{M_a d}  \right),i=1,2,\cdots,M_a
	\end{equation}
where $\hat{z}$ is the estimated root value. Then, we find the angle $\theta_{m_s}$ corresponding to maximum $P_{m_s}$, denoted by $\hat{\theta}_{M_s}$.
The correct value is solved by
	\begin{equation}\label{theta_est_final}
		\hat{\theta}_0 = \bm{\Theta_c}\left[\arg\max \left|\bm{\Theta_c}-\hat{\theta}_{M_s}\right|\right]
	\end{equation}
This method can be called as single time block root MUSIC (STB-root-MUSIC).

\section{Performance Analysis}
In this section, the performance loss and energy efficiency is analyzed for the HAD structure with mixed-ADCs.
\subsection{Performance Loss}
In this subsection, to obtain the accurate performance loss, we derive the the closed-form expressions of the CRLB in this architecture.



Consider the AB is set as $\mathbf{v}_{A,m_s}=\frac{1}{\sqrt{M_a}}\left[1~1~\cdots~1 \right]^T,~m_s=1,2,\cdots,M_s$. Thus, the covariance matrix of quantization noise can be given by
\begin{align}\label{R_wh1}
\mathbf{R}_{\mathbf{w}_q}&=\alpha\beta \mathbf{diag}(\sigma_s^2\mathbf{V}_{A,1}^H\mathbf{a}_{1}(\theta_0) \mathbf{a}_{1}^H(\theta_0)\mathbf{V}_{A,1}+\mathbf{I}_{M_1})\nonumber\\
&=\alpha\beta \mathbf{diag}\left(\frac{\sigma_s^2\|\zeta\|^2}{M_a}\mathbf{a}_{s,1}(\theta_0)\mathbf{a}_{s,1}(\theta_0)^H+\mathbf{I}_{M_1}\right)\nonumber\\
&=\alpha\beta \left(\frac{\gamma\|\zeta\|^2}{M_a}+1\right)\mathbf{I}_{M_1},
\end{align}
where $\gamma=\sigma_{s}^2 / \sigma_{n}^2=\sigma_{s}^2$ is the input SNR of the received signal,
\begin{align}\label{as1}
\mathbf{a}_{s,1}(\theta_0)=&e^{j\frac{2\pi}{\lambda}M_0d\sin\theta_0}\nonumber\\
&\times\left[1~e^{j\frac{2\pi}{\lambda}M_ad\sin\theta_0}~\cdots ~e^{j\frac{2\pi}{\lambda}(M_1-1)M_ad\sin\theta_0}\right]^T,
\end{align}
and
\begin{equation}\label{zeta}
\zeta=\sum_{m_a=1}^{M_a}e^{j\frac{2\pi}{\lambda}d_{1,m_a}\sin\theta_0}=\frac{1-e^{j\frac{2\pi}{\lambda}M_ad\sin\theta_0}}{1-e^{j\frac{2\pi}{\lambda}d\sin\theta_0}}.
\end{equation}
Therefore, we have $\mathbf{w}_q\sim\mathcal{CN}(0,\sigma_q^2\mathbf{I}_{M_1})$, where $\sigma_q^2=\alpha\beta \left(\frac{\gamma\|\zeta\|^2}{M_a}+1\right)$.
And, the FIM for the hybrid structure with mixed-ADCs is calculated in the Appendix, which is given as (\ref{Fex}). 
\begin{figure*}[hb] %
	\vspace*{4pt}
	\hrulefill
	\begin{equation}\label{Fex}
		\mathbf{F}=\left[
		\begin{array}{cc}
			\left( \frac{\xi\|\zeta\|^2}{\gamma\xi\|\zeta\|^2+M_a} \right)^2 & 0 \\
			0 & \frac{8\pi^2\gamma^2\cos^2\theta_0}{\lambda^2M_a(\gamma\xi\|\zeta\|^2+M_a)^2}\left[\|\zeta\|^4(\xi\nu-\mu^2)(\gamma\xi\|\zeta\|^2 +M_a)+M_a\xi^2\left(\|\zeta\|^2\|\Gamma\|^2-\mathbb{R}\left[\left(\Gamma^H\zeta\right)^2\right]\right)\right]
		\end{array}
		\right]
	\end{equation}
\end{figure*}
Due to
\begin{equation}\label{CRLBdefine}
\mathbf{CRLB}=\frac{1}{N}\mathbf{F}^{-1},
\end{equation}
the CRLB of the direction can be expressed as
\begin{equation}\label{vartheta}
\emph{var}(\hat{\theta_0})\geq\frac{\lambda^2M_a(\gamma\xi\|\zeta\|^2+M_a)^2 } {8\varphi N\pi^2\gamma^2\cos^2\theta_0},
\end{equation}
where
\begin{align}\label{xiaoA}
\varphi=&\|\zeta\|^4(\xi\nu-\mu^2)(\gamma\xi\|\zeta\|^2 +M_a)\nonumber\\
&+M_a\xi^2\left(\|\zeta\|^2\|\Gamma\|^2-\mathbb{R}\left[\left(\Gamma^H\zeta\right)^2\right]\right).
\end{align}
Referring to \cite{Shiarxiv}, we define the performance loss factor as
\begin{equation}\label{plossh}
\eta_{PL}=\frac{\emph{CRLB}_{\theta_0}}{\emph{CRLB}_{\theta_0}^{M_a=1,\kappa=1}}.
\end{equation}
From (\ref{vartheta}), the performance loss factor can be given as (\ref{plosscf}).
\begin{figure*}[hb] 
	\vspace*{5pt}
	\hrulefill
	\begin{equation}\label{plosscf}
		\eta_{PL}=\frac{M^2 d^2(M^2-1)(\gamma\xi\|\zeta\|^2+M_a)^2}{12(\gamma M+1)\left[\|\zeta\|^4(\xi\nu-\mu^2)(\gamma\xi\|\zeta\|^2 +M_a)+M_a\xi^2\left(\|\zeta\|^2\|\Gamma\|^2-\mathbb{R}\left[\left(\Gamma^H\zeta\right)^2\right]\right)\right]}.
	\end{equation}
\end{figure*}
Different from the performance loss factor in \cite{shiDOAmixedADC}, $M_a/M=1/M_s\approx0$ is invalid in HAD structure, especially when $\gamma$ is low and $M_a$ is high. Thus, the impact of the $M$ on the performance loss is hard to be ignored in our architecture.

\subsection{Energy Efficiency}
Although we have investigated the performance loss, it is hard to assess the the HAD structure with mixed-ADCs if we only consider the performance loss. Therefor, to achieve a better trade-off, it is obliged to discussed the power consumption of the HAD structure with mixed-ADCs. Due to this reason, the power consumption and energy efficiency for the DOA estimation is studied in the HAD structure with mixed-ADCs.

According to \cite{shiDOAmixedADC}, the energy efficiency factor can be defined as
\begin{equation}\label{etaee}
\eta_{EE}=\frac{CRLB_{\theta_0}^{-\frac{1}{2}}}{P_{t}}~~1/\mathrm{degree}/\mathrm{W},
\end{equation}
where $P_{t}$ is the total power consumption in the system \cite{Zhangtcom}, which can be given by
\begin{align}\label{Pall}
P_{t}&=MP_{APS}+P_{RF}+M_0(P_{AGC}+P_{HADC})\nonumber\\
&~~~+M_1(\chi P_{AGC}+P_{LADC})
\end{align}
where $P_{APS}$, $P_{RF}$, $P_{AGC}$, $P_{HADC}$ and $P_{LADC}$ denotes the the power consumption of the analog phase shifter, the RF chain, the automatic gain control (AGC), the high-resolution ADC and the low-resolution ADC, respectively.
Among them,
\begin{equation}\label{PRF}
P_{RF}=M_s(P_{LNA}+P_{m}+P_{f}+P_{IFA})+P_{fsyc}
\end{equation}
where $P_{LNA}$, $P_{m}$, $P_{f}$, $P_{IFA}$ and$P_{fsyc}$ represent the power consumption of the low-noise amplifiers (LNA), the mixer, the filter, the intermediate frequency amplifier (IFA) and the frequency synthesizer, respectively.
Besides, $\chi$ is a flag function connected with the number of the ADCs' quantization bit, which is given by
\begin{equation}\label{flagc}
\chi=\left\{
\begin{array}{cc}
  0,~~ & b=1, \\
  1,~~ & b>1.
\end{array}
\right.
\end{equation}
Furthermore, by consulting \cite{CuiEnergy}, $P_{ADC}$ can be expressed as
\begin{equation}\label{PADC}
P_{ADC}\approx\frac{3V_{dd}^2L_{min}(f_{cor}+2B)}{10^{-0.1525b+4.838}}.
\end{equation}
where $V_{dd}$ is the power supply of converter, $B$ is the bandwidth of received signal, $f_{cor}$ is the corner frequency of the 1/f noise \cite{CuiEnergy}, and $L_{min}$ denotes the the minimum channel length for the given CMOS technology. It is difficult to analyse the (\ref{etaee}). Thus, detailed results are presented in Section IV.

%
%

\section{Simulation and Discussion}
In this section, simulation results are presented to verify the the accuracy of the analytical results. Then, the curves of performance loss for different conditions are shown. Then, the root-MUSIC-based algorithm in the HAD architecture is demonstrated to show the practical performance of this architecture. Moreover, the energy efficiency is investigated to present the advantage of the HAD architecture with mixed-ADCs and provide the suggestions of choices for quantization bits.

In the simulations, we assume that the parameters are chosen as follow: $\theta_0=15^\circ$, $N=32$ $\kappa=1/4$ and $M=128$.
To assess the performance of the methods in practical, the root mean square error (RMSE) is adopted in simulations, which is given by
\begin{equation}\label{RMSE}
RMSE=\sqrt{\frac{1}{N_t}\sum\limits_{n_t=1}^{N_t}(\hat{\theta}_{n_t} - \theta_{0})^2},
\end{equation}
where $N_t$ is the number of numerical simulations, which is set as 18000.

\begin{figure} [t!]
	\centering
	\subfloat[ $\theta_0=23^\circ$\label{1a}]{
		\includegraphics[width=0.48\textwidth]{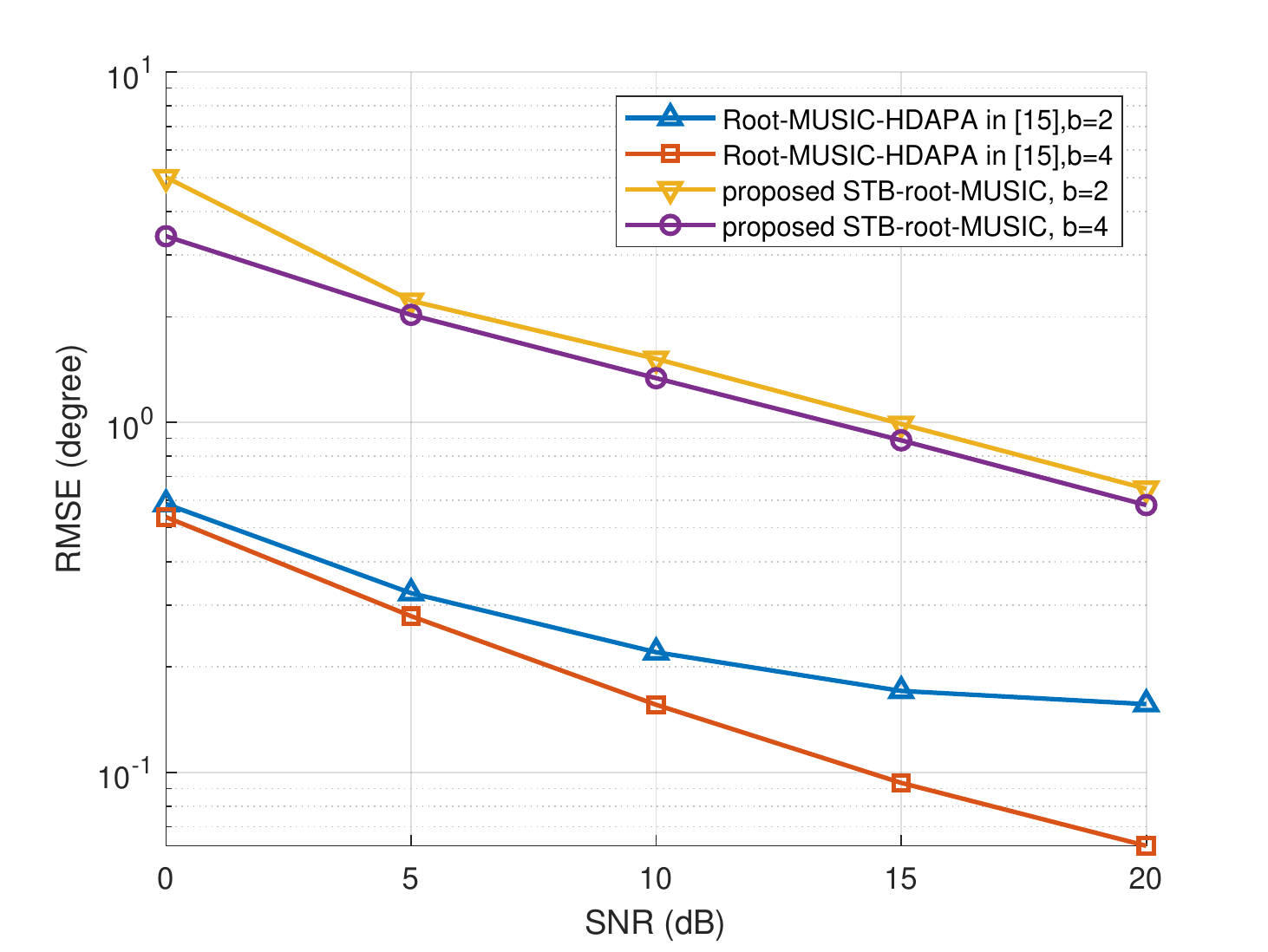}}
	\quad
	\subfloat[$\theta_0=73^\circ$\label{1b}]{
		\includegraphics[width=0.48\textwidth]{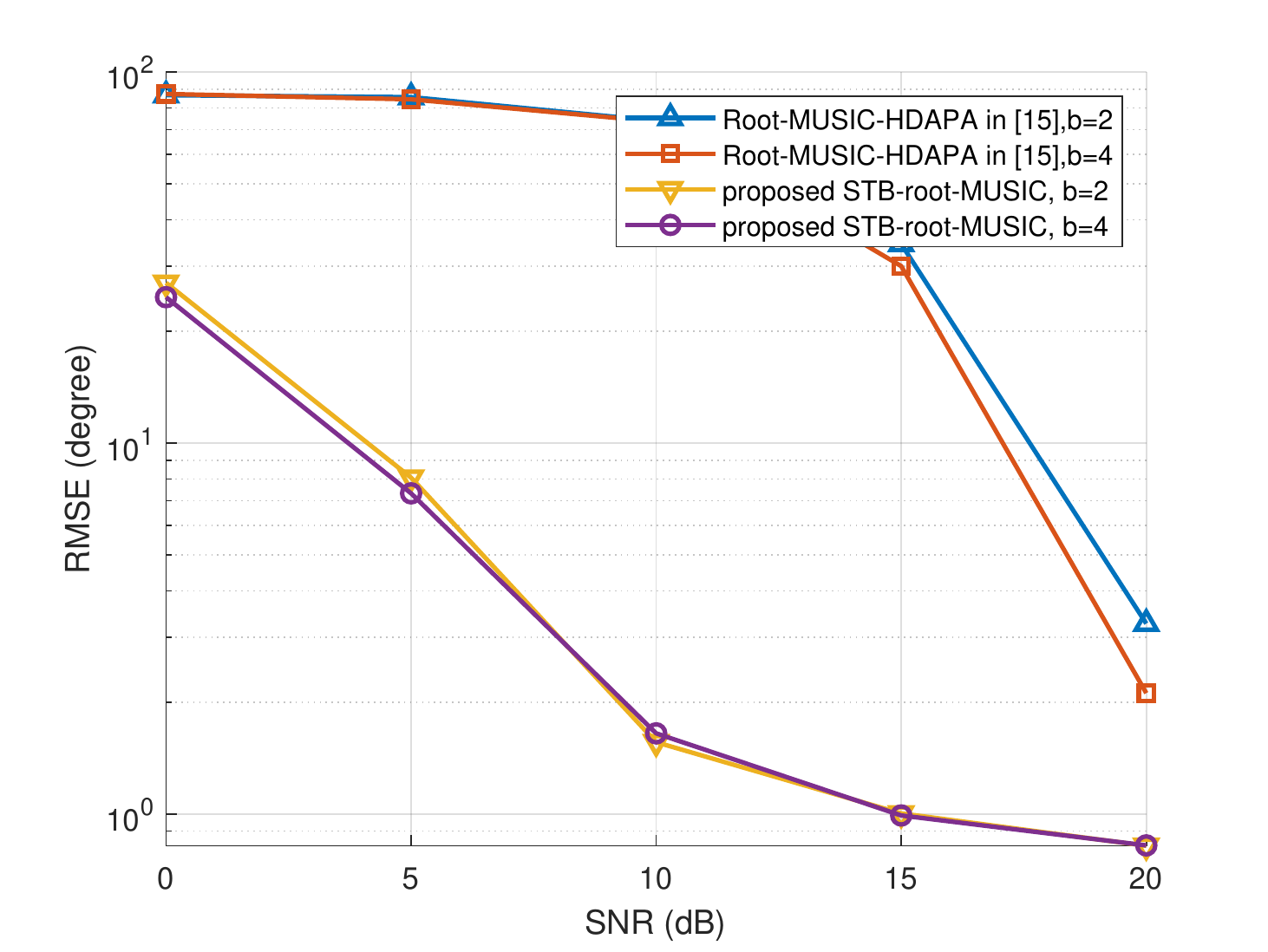}}
	\caption{RMSE over SNR with different methods}
	\label{fig_RMSE_SNR_comp} 
\end{figure}

In Fig.~\ref{fig_RMSE_SNR_comp}, the RMSE over SNR with different methods is presented. Two cases are considered: (a) $\theta_0=23^\circ$ and (b) $\theta_0=73^\circ$. From Figure.~\ref{fig_RMSE_SNR_comp} (a), we can find that when the direction is in the beamwidth of Root-MUSIC-HDAPA, the performance of Root-MUSIC-HDAPA in \cite{shuDOA} is better than proposed method. 
However, observing Figure.~\ref{fig_RMSE_SNR_comp} (b), the performance is reversed when the direction is out of beamwidth. Note that, as $M_a$ increases, beamwidth becomes small and robustness of Root-MUSIC-HDAPA will be worse. 
Additionally, the proposed method is much more robust to different directions due to that the performance of proposed method is similar in two cases. 

\begin{figure}
  \centering
  \includegraphics[width=0.5\textwidth]{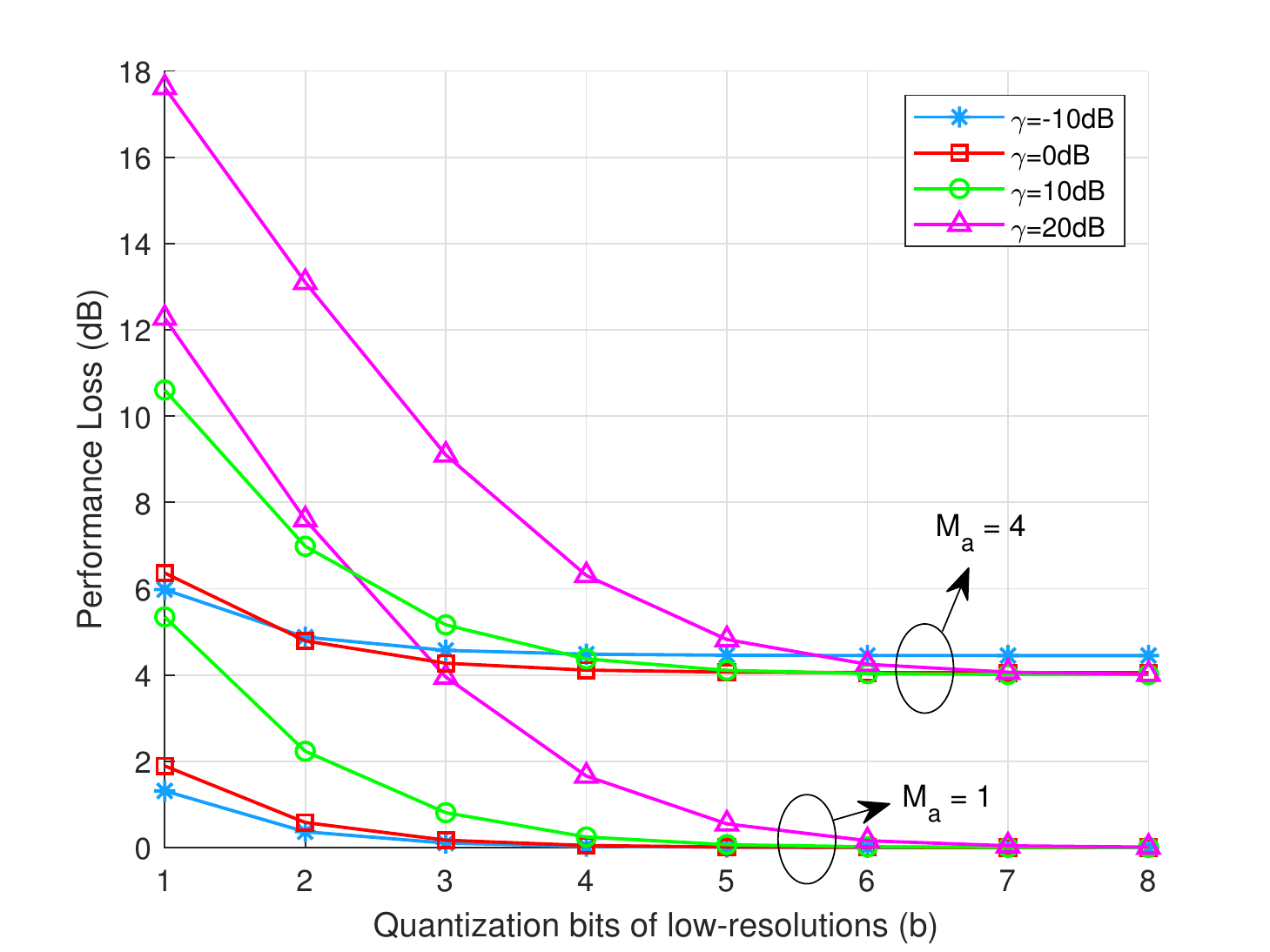}\\
  \caption{Performance loss over quantization bits of low-resolution ADCs.}\label{fig_PL_beta}
\end{figure}
Fig.~\ref{fig_PL_beta} illustrates the performance loss versus the quantization bits for different $M_a$. It is clear that the performance loss decreases as the quantization bits increase for all cases. And, the performance loss is bigger at higher SNR. We also find that increasing the quantization bits has negligible performance improvement when $b>5$. It is seen from Fig.~\ref{fig_PL_beta} that the performance loss for $\gamma=-10dB$ is worse than other cases as the quantization bits increase when $M_a=4$. In conclusion, just 2-3 bits of the resolution can causes little performance loss. In addition, at the high SNR, adopting 4-5 bits' ADCs is a better choice. However, bigger quantization bits is not recommended.

\begin{figure}
  \centering
  \includegraphics[width=0.5\textwidth]{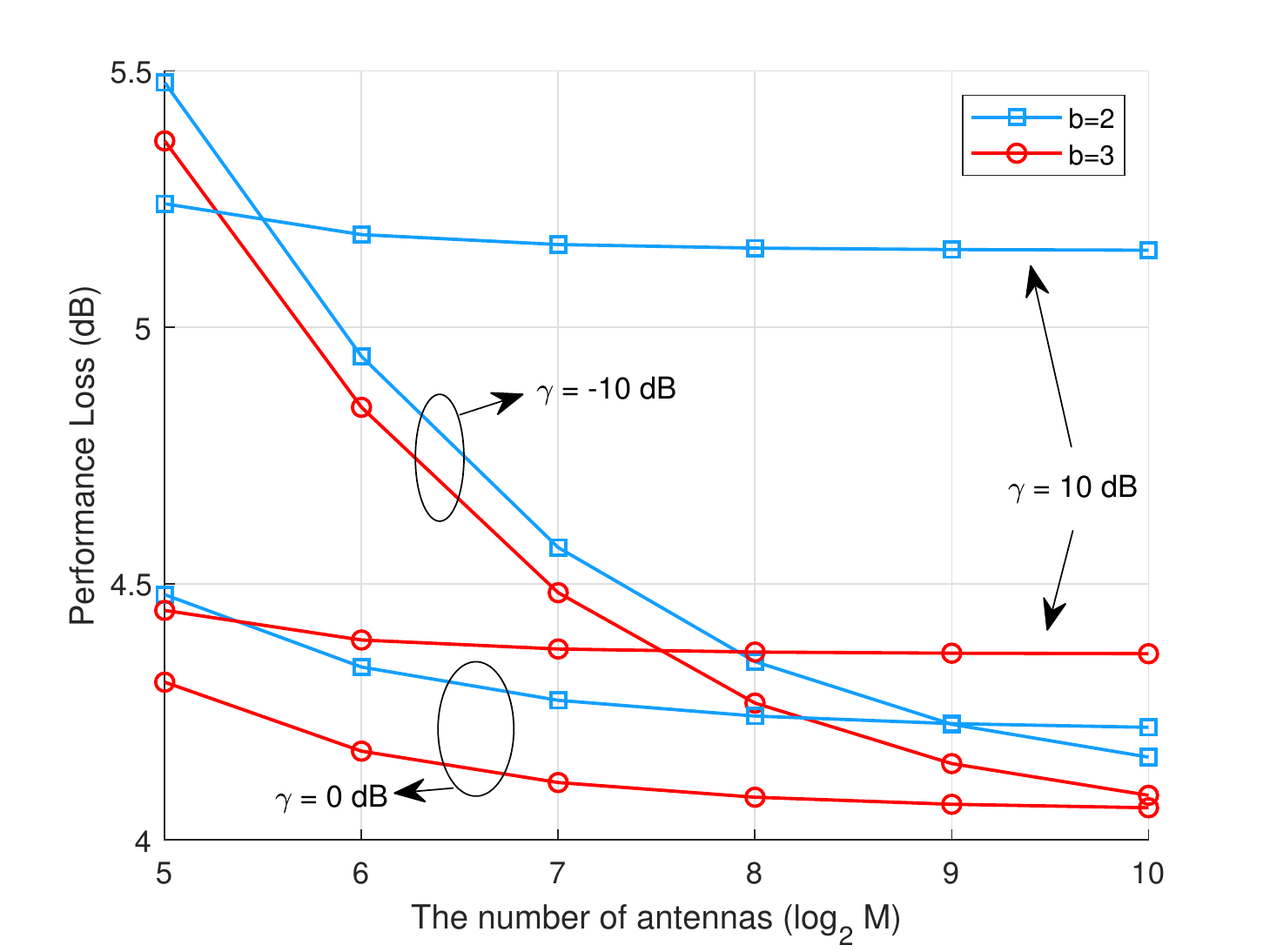}\\
  \caption{Performance loss over the number of antennas with different quantization bits and SNR. }\label{fig_PL_M}
\end{figure}
In Fig.~\ref{fig_PL_M}, the performance loss for the HAD architecture with mixed-ADCs against the number of antennas is plotted. The number of antennas $M$ ranges from 32 to 1024. And, the quantization bits of two different values, $b=2,3$, are considered. It can be seen that all curves decrease as the number of antennas increases. However, as the SNR increases, the slope of the corresponding curves becomes smaller. Especially, the curves are almost flat when $\gamma=10dB$. An insightful observation is that the HAD structure with mixed-ADCs is very suitable to many future practical application scenarios because the number of antennas in the massive MIMO system is going to increase  bigger and bigger and the SNR in many DOA estimation scenarios is low.
%

\begin{figure}
  \centering
  \includegraphics[width=0.5\textwidth]{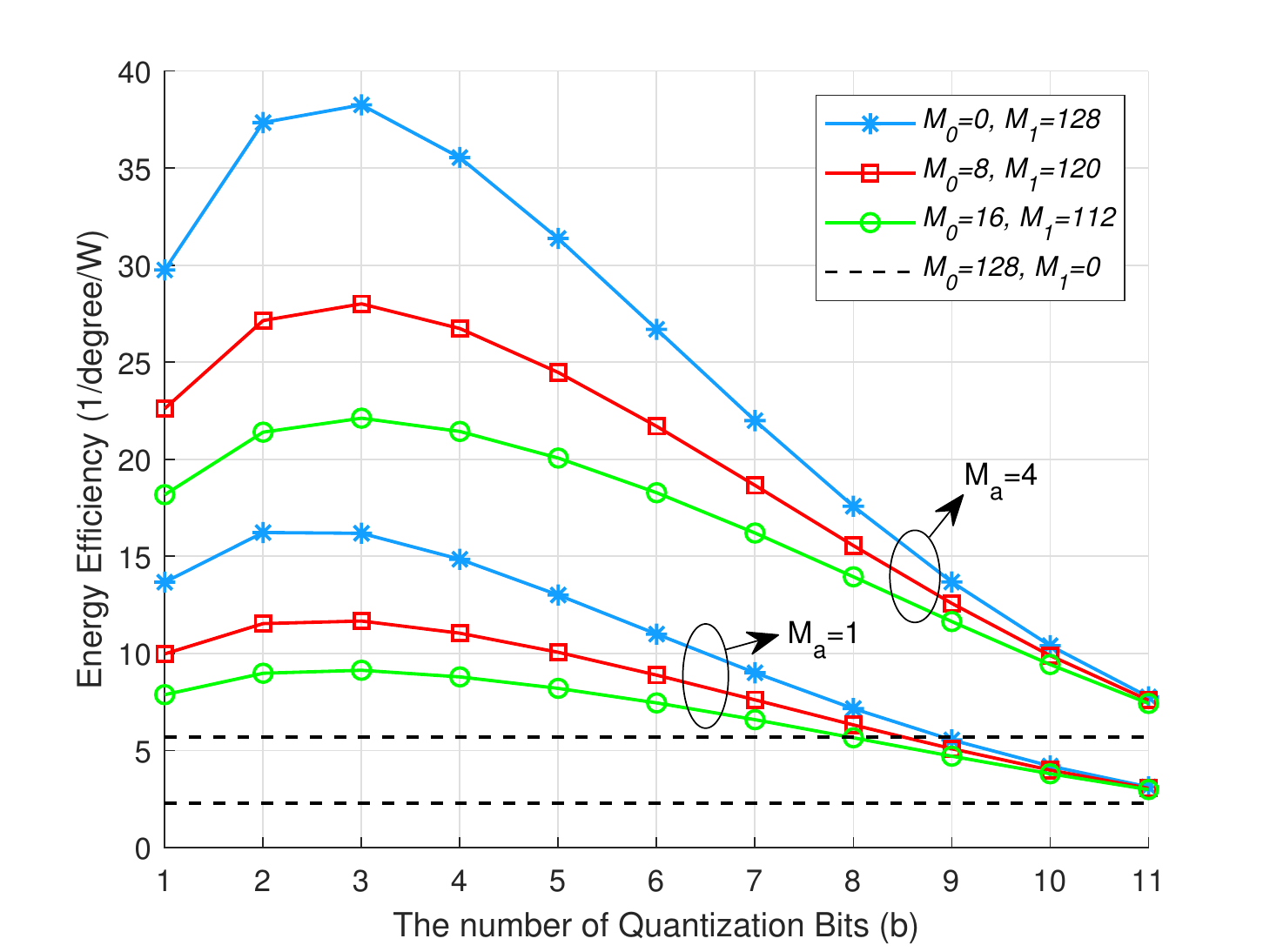}\\
  \caption{Energy efficiency of the HAD architecture with mixed-ADCs against different quantization bits}\label{fig_EE}
\end{figure}
In Fig.~\ref{fig_EE}, the energy efficiency for the HAD architecture with mixed-ADCs versus the ADCs' quantization bits is shown. Herein, three different cases are considered as follow: $1)M_0=0,~M_1=128$, $2)M_0=8,~M_1=120$, $3)M_0=16,~M_1=112$.
In our simulation, the values in massive MIMO systems are chosen as follow: $P_{APS}=1~\textrm{mW}$, $P_{LNA}=20~\textrm{mW}$, $P_{mix}=30.3~\textrm{mW}$, $P_{fil}=2.5~\textrm{mW}$, $P_{IFA}=3~\textrm{mW}$, $P_{syc}=50.5~\textrm{mW}$, $P_{AGC}=2~\textrm{mW}$, $V_{dd}=3~\textrm{V}$, $B=20~\textrm{MHz}$, $L_{min}=0.5~\mu \textrm{m}$ and $f_{cor}=1~\textrm{MHz}$ as in \cite{Zhangtcom,GaoHAD,HeEnergy}. And, 12 is adopted as the number of quantization bit for high-resolution ADCs. As can be seen, all curves have peaks and the positions of peaks. And, compared with curves with $M_a=1$, the positions of peaks moves to right for the curves with $M_a=4$. This important finding demonstrates that compared with the full digital structure with mixed-ADCs, the HAD structure with mixed-ADCs has much higher energy efficiency. In addition, the energy efficiency decreases as the proportion of the high-resolution ADCs increases. Thus, the array with pure high-resolution ADCs has lowest energy efficiency, which implies that adopting low-resolution ADCs is effective. In summary, the HAD architecture with mixed-ADCs is superior to the HAD architecture and the mixed-ADC structure, which has higher energy efficiency.

\begin{figure}
  \centering
  \includegraphics[width=0.5\textwidth]{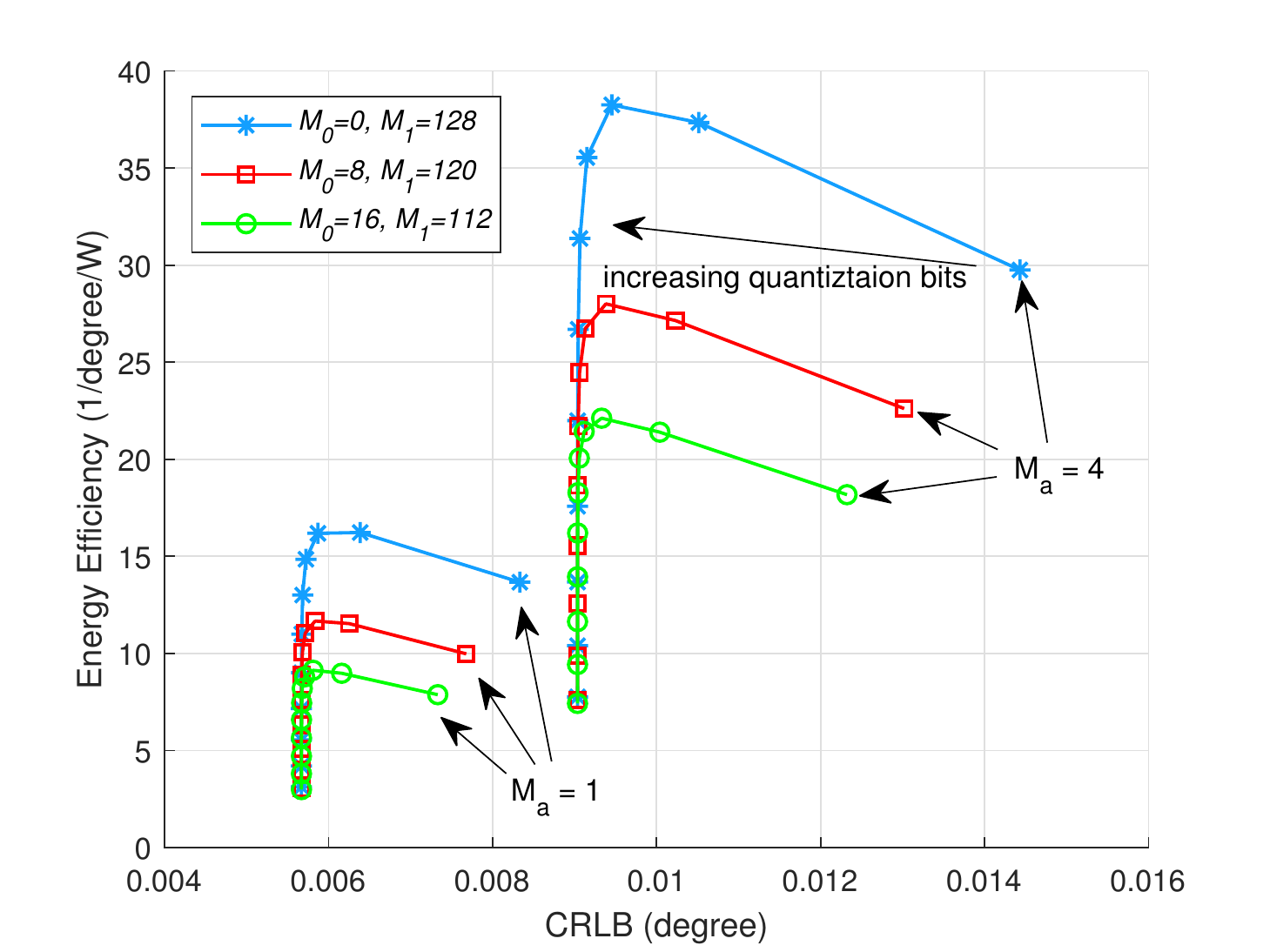}\\
  \caption{Trade-off between the CRLB and the energy efficiency}\label{fig_EE_CRLB}
\end{figure}
In Fig.~\ref{fig_EE_CRLB}, we show the trade-off between the CRLB and energy efficiency with different proportions of the high-resolution ADCs. The curves of the full digital structure with mixed-ADCs ($M_a=1$) are also plotted as a benchmark. It is seen that the HAD structure with mixed-ADCs has much higher energy efficiency. And, adopting pure low-resolution ADCs can achieve the best energy efficiency. However, considering the poor performance of DOA estimation and low achievable rate in \cite{Zhangjsac}, it is hard to employ the pure low-resolution ADCs structure in the practical massive MIMO systems.
Moreover, we can conclude that the energy efficiency increases as the proportion of the high-resolution ADCs, $\kappa$, decreases. When we increase the number of quantization bits from 1 to 4, the CRLB will decrease quickly. While the performance almost has no improvement when quantization bits range from 4 to 12 bits. In addition, increasing ADCs' resolution in the HAD structure has more performance improvement than that in the full digital structure.

\section{Conclusion}
In this paper, the performance of the DOA estimation for the hybrid analog and digital structure with mixed-ADCs  has been investigated. We have established the system model of the DOA estimation for the HAD massive MIMO system with mixed-ADCs. A novel DOA method is proposed to estimate the direction in single time block. 
Then, the performance loss factor has been defined for that architecture. Furthermore, the closed-form expression of the corresponding CRLB has been derived. Based on that, the closed-form expression of performance loss factor has been also derived. In addition, we have studyed the energy efficiency for that architecture and found that adopting 2-4 quantization bit for the low-resolution ADCs can obtain a better trade-off. Finally, we can draw the conclusion that the HAD structure with mixed-ADCs can obtain a much lower power consumption and radio frequency cost with little performance loss.

\appendix[Derivation of the FIM for the HAD Architecture with Mixed-ADCs]

In this section, to have access to the Cram\'{e}r-Rao lower bound (CRLB) for hybrid structure with mixed-ADCs, we derive the closed-form expression fo the corresponding Fisher information matrix (FIM). Considering the $\theta_0$ and $\gamma$ are all unknown, $\mathbf{F}$ can be divided as
\begin{equation}\label{Fall}
\mathbf{F} = \left[
\begin{array}{cc}
  \mathbf{F}_{\gamma,\gamma} & \mathbf{F}_{\gamma,\theta_0} \\
  \mathbf{F}_{\gamma,\theta_0} & \mathbf{F}_{\theta_0,\theta_0}
\end{array}
\right].
\end{equation}
In accordance with \cite{DOA}, the elements of $\mathbf{F}$ can be calculated by
\begin{equation}\label{Fgammagamma}
\mathbf{F}_{\gamma,\gamma} =\mathbf{Tr}\left\{ \mathbf{R}_{\mathbf{y} }^{-1} \frac{\partial \mathbf{R}_{\mathbf{y} }}{\partial \gamma} \mathbf{R}_{\mathbf{y} }^{-1} \frac{\partial \mathbf{R}_{\mathbf{y} }}{\partial \gamma} \right\},
\end{equation}
\begin{equation}\label{Fgammatheta}
\mathbf{F}_{\gamma,\theta_0} =\mathbf{Tr}\left\{ \mathbf{R}_{\mathbf{y} }^{-1} \frac{\partial \mathbf{R}_{\mathbf{y} }}{\partial \gamma} \mathbf{R}_{\mathbf{y} }^{-1} \frac{\partial \mathbf{R}_{\mathbf{y} }}{\partial \theta_0} \right\},
\end{equation}
\begin{equation}\label{Fthetagamma}
\mathbf{F}_{\theta_0,\gamma} =\mathbf{Tr}\left\{ \mathbf{R}_{\mathbf{y} }^{-1} \frac{\partial \mathbf{R}_{\mathbf{y} }}{\partial \theta_0} \mathbf{R}_{\mathbf{y} }^{-1} \frac{\partial \mathbf{R}_{\mathbf{y} }}{\partial \gamma} \right\},
\end{equation}
and
\begin{equation}\label{Fthetatheta}
\mathbf{F}_{\theta_0,\theta_0} =\mathbf{Tr}\left\{ \mathbf{R}_{\mathbf{y} }^{-1} \frac{\partial \mathbf{R}_{\mathbf{y} }}{\partial \theta_0} \mathbf{R}_{\mathbf{y} }^{-1} \frac{\partial \mathbf{R}_{\mathbf{y} }}{\partial \theta_0} \right\}.
\end{equation}

Firstly, let us abbreviate $s(n)$, $\mathbf{y}(n)$, $\mathbf{a}(\theta_{0})$ and $\mathbf{w}(n)$ as $s$, $\mathbf{y}$, $\mathbf{a}$ and $\mathbf{w}$, respectively.
Then, $\mathbf{y} $ can be rewritten as
\begin{equation}\label{yh_T}
\mathbf{y} =\mathbf{T}\mathbf{V}_A^H\mathbf{a}s+\mathbf{T}\mathbf{w}+\mathbf{q} ,
\end{equation}
where
\begin{equation}\label{T}
\mathbf{T}=\left[
\begin{array}{cc}
  \mathbf{I}_{M_0}              & \mathbf{0}_{M_0 \times M_1} \\
  \mathbf{0}_{M_1 \times M_0}   & \alpha\mathbf{I}_{M_1}
\end{array}
\right]
\end{equation}
and
\begin{equation}\label{q}
\mathbf{q}=\left[
\begin{array}{c}
    \mathbf{0}_{M_0 \times 1} \\
    \mathbf{w}_{q}
\end{array}
\right].
\end{equation}
Thus, the covariance matrix of $\mathbf{y} $ is derived by
\begin{equation}\label{Ryh}
\mathbf{R}_{\mathbf{y}} =\mathbb{E}[\mathbf{y}\mathbf{y}^H]=\gamma\mathbf{T}\mathbf{V}_A^H\mathbf{a}\mathbf{a}^H\mathbf{V}_A\mathbf{T}^H+\mathbf{Q},
\end{equation}
where
\begin{align}\label{Qh}
\mathbf{Q} &=\mathbb{E}\left[\mathbf{T}\mathbf{w}\mathbf{w}^H\mathbf{T}^H+\mathbf{q}\mathbf{q}^H\right]\nonumber\\
&=\left[
\begin{array}{cc}
  \mathbf{I}_{M_0}              & \mathbf{0}_{M_0 \times M_1} \\
  \mathbf{0}_{M_1 \times M_0}   & \left(\alpha^2+\sigma_q^2\right)\mathbf{I}_{M_1}
\end{array}
\right].
\end{align}
where $\sigma_q^2=\alpha\beta \left(\frac{\gamma\|\zeta\|^2}{M_a}+1\right)$ is defined in (\ref{R_wh1}).
Accordingly, the partial derivatives of $\mathbf{R}_{\mathbf{y}}$ are given by
\begin{equation}\label{dRydgamma}
\frac{\partial\mathbf{R}_{\mathbf{y} }}{\partial\gamma}=\mathbf{T}\mathbf{V}_A^H\mathbf{a}\mathbf{a}^H\mathbf{V}_A\mathbf{T}^H
\end{equation}
and
\begin{equation}\label{dRyh}
\frac{\partial\mathbf{R}_{\mathbf{y} }}{\partial\theta_0}=\gamma \mathbf{T}\mathbf{V}_A^H(\mathbf{a}'\mathbf{a}^H+\mathbf{a}\mathbf{a}'^H)\mathbf{V}_A\mathbf{T}^H.
\end{equation}
where $\mathbf{a}'$ is the derivative for the array manifold to $\theta_0$, which is given by
\begin{equation}\label{da}
\mathbf{a}'=\frac{d}{d\theta_0}\mathbf{a}(\theta_0)=j\frac{2\pi}{\lambda}\cos\theta_0\mathbf{D}\mathbf{a}
\end{equation}
where
\begin{equation}\label{D}
\mathbf{D}=\left[
\begin{array}{cccc}
d_1     & 0         & \cdots & 0 \\
0       & d_2       & \cdots & 0 \\
\vdots  & \vdots    & \ddots & \vdots \\
0       & 0         & \cdots & d_M
\end{array}
\right].
\end{equation}
By utilizing the well-known Sherman-Morrison matrix identity in \cite{Matrix},
\begin{equation}\label{Sherman}
(A+BCD)^{-1}=A^{-1}-A^{-1}B(C^{-1}+DA^{-1}B)^{-1}DA^{-1},
\end{equation}
we can obtain
\begin{equation}\label{Ryh_inv}
\mathbf{R}_{\mathbf{y} }^{-1}=\mathbf{Q} ^{-1}-\frac{\mathbf{Q} ^{-1}\mathbf{T}\mathbf{V}_A ^H\mathbf{a}\mathbf{a}^H\mathbf{V}_A\mathbf{T}^H\mathbf{Q} ^{-1}}{\gamma^{-1}+ \mathbf{a}^{H}\mathbf{V}_A\mathbf{T}^H\mathbf{Q} ^{-1}\mathbf{T}\mathbf{V}_A^H\mathbf{a}}.
\end{equation}

Let us derive the $\mathbf{F}_{\gamma,\gamma}$ firstly, which can be expanded to
\begin{align}\label{Fgammagammaex}
\mathbf{F}_{\gamma,\gamma} &=\mathbf{Tr}\left\{ \mathbf{R}_{\mathbf{y} }^{-1} \mathbf{T}\mathbf{V}_A^H\mathbf{a}\mathbf{a}^H\mathbf{V}_A\mathbf{T}^H \mathbf{R}_{\mathbf{y} }^{-1} \mathbf{T}\mathbf{V}_A^H\mathbf{a}\mathbf{a}^H\mathbf{V}_A\mathbf{T}^H\right\}\nonumber\\
&=\left[\mathbf{a}^H\mathbf{V}_A\mathbf{T}^H\left(\mathbf{Q} ^{-1}\frac{}{} \right.\right.\nonumber\\
&~~~\left.\left.-\frac{\mathbf{Q} ^{-1}\mathbf{T}\mathbf{V}_A ^H\mathbf{a}\mathbf{a}^H\mathbf{V}_A\mathbf{T}^H\mathbf{Q} ^{-1}}{\gamma^{-1}+ \mathbf{a}^{H}\mathbf{V}_A\mathbf{T}^H\mathbf{Q} ^{-1}\mathbf{T}\mathbf{V}_A^H\mathbf{a}}\right)\mathbf{T}\mathbf{V}_A^H\mathbf{a}\right]^2\nonumber\\
&=\left[\mathbf{a}^H\mathbf{B}\mathbf{a}-\frac{(\mathbf{a}^H\mathbf{B}\mathbf{a})^2}{\gamma^{-1}+\mathbf{a}^H\mathbf{B}\mathbf{a}}\right]^2,
\end{align}
where
\begin{equation}\label{B}
\mathbf{B}=\mathbf{V}_A\mathbf{T}^H\mathbf{Q} ^{-1}\mathbf{T}\mathbf{V}_A^H .
\end{equation}
By resorting to the Kronecker product, we have
\begin{equation}\label{Bkp}
\mathbf{B}=\frac{1}{M_a}\mathbf{C}\otimes\left(\mathbf{1}_{M_a}\mathbf{1}_{M_a}^H\right),
\end{equation}
where
\begin{equation}\label{C}
\mathbf{C}=\left[
\begin{array}{cc}
  \mathbf{I}_{M_0}              & \mathbf{0}_{M_0 \times M_1} \\
  \mathbf{0}_{M_1 \times M_0}   & \frac{\alpha^2}{\alpha^2+\sigma_q^2}\mathbf{I}_{M_1}
\end{array}
\right],
\end{equation}
and
\begin{equation}\label{akp}
\mathbf{a}=\mathbf{a}_{s}\otimes \mathbf{a}_{a},
\end{equation}
where
\begin{equation}\label{as}
\mathbf{a}_s=[1,e^{j\frac{2\pi}{\lambda}M_ad\sin\theta_0},\cdots,e^{j\frac{2\pi}{\lambda}(M_s-1)M_ad\sin\theta_0)}]^T,
\end{equation}
and
\begin{equation}\label{aa}
\mathbf{a}_a=[1,e^{j\frac{2\pi}{\lambda}d\sin\theta_0},\cdots,e^{j\frac{2\pi}{\lambda}(M_a-1)d\sin\theta_0}]^T.
\end{equation}
Hence, referring to the properties of Kronecker product in \cite{Matrix},
we have
\begin{align}\label{aba}
\mathbf{a}^{H}\mathbf{B}\mathbf{a}&=\frac{1}{M_a}(\mathbf{a}_s\otimes \mathbf{a}_a)^H(\mathbf{C}\otimes\mathbf{1}_{M_a}\mathbf{1}_{M_a}^H)(\mathbf{a}_s\otimes \mathbf{a}_a) \nonumber\\
&=\frac{1}{M_a}(\mathbf{a}_s^H\mathbf{C}\mathbf{a}_s\otimes \mathbf{a}_a^H\mathbf{1}_{M_a}\mathbf{1}_{M_a}^H\mathbf{a}_a ) \nonumber\\
&=\frac{1}{M_a}\left( M_0 + \frac{\alpha^2}{\alpha^2+\sigma_q^2}M_1 \right) \|\mathbf{1}_{M_a}^H\mathbf{a}_a\|^2 \nonumber\\
&=\frac{1}{M_a}\xi\|\zeta\|^2,
\end{align}
where
\begin{equation}\label{xih}
\xi=M_0 + \frac{\alpha^2}{\alpha^2+\sigma_q^2}M_1.
\end{equation}
Thus, $\mathbf{F}_{\gamma,\gamma}$ can be expressed as
\begin{equation}\label{Fgammagammaend}
\mathbf{F}_{\gamma,\gamma}=\left( \frac{\xi\|\zeta\|^2}{\gamma\xi\|\zeta\|^2+M_a} \right)^2.
\end{equation}
Similarly, $\mathbf{F}_{\gamma,\theta_0}$ can be given by
\begin{align}\label{Fgammathetaex}
\mathbf{F}_{\gamma,\theta_0} &=\gamma \mathbf{Tr}\left\{ \mathbf{R}_{\mathbf{y} }^{-1} \mathbf{T}\mathbf{V}_A^H\mathbf{a}\mathbf{a}^H\mathbf{V}_A\mathbf{T}^H \mathbf{R}_{\mathbf{y} }^{-1} \mathbf{T}\mathbf{V}_A^H(\mathbf{a}'\mathbf{a}^H\right. \nonumber\\
&~~~\left.+\mathbf{a}\mathbf{a}'^H)\mathbf{V}_A\mathbf{T}^H\right\}\nonumber\\
&=\gamma\mathbf{a}^H\mathbf{V}_A\mathbf{T}^H \mathbf{R}_{\mathbf{y}}^{-1} \mathbf{T}\mathbf{V}_A^H\mathbf{a}\left(\mathbf{a}^H\mathbf{V}_A\mathbf{T}^H \mathbf{R}_{\mathbf{y}}^{-1} \mathbf{T}\mathbf{V}_A^H\mathbf{a}'\right.\nonumber\\
&~~~\left.+\mathbf{a}'^H\mathbf{V}_A\mathbf{T}^H \mathbf{R}_{\mathbf{y} }^{-1} \mathbf{T}\mathbf{V}_A^H\mathbf{a}\right).
\end{align}
And, the elements in the bracket in (\ref{Fgammathetaex}) can be simplified as
\begin{align}\label{Fasub}
\mathbf{a}^H&\mathbf{V}_{\mathbf{A}}\mathbf{T}^H\mathbf{R}_{\mathbf{y} }^{-1} \mathbf{T}\mathbf{V}_{\mathbf{A}}^H\mathbf{a}'\nonumber\\
&=\mathbf{a}^H\mathbf{V}_{\mathbf{A}}\mathbf{T}^H \left(\mathbf{Q} ^{-1} -\frac{\mathbf{Q} ^{-1}\mathbf{T}\mathbf{V}_A ^H\mathbf{a}\mathbf{a}^H\mathbf{V}_A\mathbf{T}^H\mathbf{Q} ^{-1}}{\gamma^{-1}+ \mathbf{a}^{H}\mathbf{B}\mathbf{a}} \right) \nonumber\\
&~~~\times \mathbf{T}\mathbf{V}_{\mathbf{A}}^H\mathbf{a}'\nonumber\\
&=j\frac{2\pi}{\lambda}\cos\theta_0\left(1-\frac{\mathbf{a}^H\mathbf{B}\mathbf{a}}{\gamma^{-1}+\mathbf{a}^H\mathbf{B}\mathbf{a}}\right) \mathbf{a}^H\mathbf{B}\mathbf{D}\mathbf{a}.
\end{align}
and
\begin{align}\label{Fcsub}
\mathbf{a}'^H&\mathbf{V}_{\mathbf{A}}\mathbf{T}^H\mathbf{R}_{\mathbf{y} }^{-1} \mathbf{T}\mathbf{V}_{\mathbf{A}}^H\mathbf{a}\nonumber\\
&=\mathbf{a}'^H\mathbf{V}_{\mathbf{A}}\mathbf{T}^H \left(\mathbf{Q} ^{-1} -\frac{\mathbf{Q} ^{-1}\mathbf{T}\mathbf{V}_A ^H\mathbf{a}\mathbf{a}^H\mathbf{V}_A\mathbf{T}^H\mathbf{Q} ^{-1}}{\gamma^{-1}+ \mathbf{a}^{H}\mathbf{B}\mathbf{a}} \right) \nonumber\\
&~~~\times \mathbf{T}\mathbf{V}_{\mathbf{A}}^H\mathbf{a}\nonumber\\
&=-j\frac{2\pi}{\lambda}\cos\theta_0\left(1-\frac{\mathbf{a}\mathbf{B}\mathbf{a}}{\gamma^{-1}+\mathbf{a}\mathbf{B}\mathbf{a}}\right) \mathbf{a}^H\mathbf{D}\mathbf{B}\mathbf{a},
\end{align}
respectively. Obviously, $\mathbf{B}\mathbf{D}=\mathbf{D}\mathbf{B}$, hence we can conclude that $\mathbf{F}_{\gamma,\theta_0}=0$.
In addition,
\begin{align}\label{Fthetagammaex}
\mathbf{F}_{\theta_0,\gamma} &=\mathbf{Tr}\left\{ \mathbf{R}_{\mathbf{y} }^{-1} \frac{\partial \mathbf{R}_{\mathbf{y} }}{\partial \theta_0} \mathbf{R}_{\mathbf{y} }^{-1} \frac{\partial \mathbf{R}_{\mathbf{y} }}{\partial \gamma} \right\}\nonumber\\
&=\mathbf{Tr}\left\{ \mathbf{R}_{\mathbf{y} }^{-1} \frac{\partial \mathbf{R}_{\mathbf{y} }}{\partial \gamma} \mathbf{R}_{\mathbf{y} }^{-1} \frac{\partial \mathbf{R}_{\mathbf{y} }}{\partial \theta_0} \right\}\nonumber\\
&=\mathbf{F}_{\gamma,\theta_0}=0.
\end{align}
Finally, the $\mathbf{F}_{\theta_0,\theta_0}$ can be written as (\ref{Fhe}), 
\begin{figure*}[hb] 
	\vspace*{4pt}
	\hrulefill
	\begin{align}\label{Fhe}
		\mathbf{F}_{\theta_0,\theta_0}&=\gamma^2\mathbf{Tr}\left\{\mathbf{R}_{\mathbf{y}}^{-1} \mathbf{T}\mathbf{V}_{\mathbf{A}}^H(\dot{\mathbf{a}}\mathbf{a}^H+\mathbf{a}\dot{\mathbf{a}}^H)\mathbf{V}_{\mathbf{A}}\mathbf{T}^H\mathbf{R}_{\mathbf{y}}^{-1} \mathbf{T}\mathbf{V}_{\mathbf{A}}^H(\dot{\mathbf{a}}\mathbf{a}^H+\mathbf{a}\dot{\mathbf{a}}^H)\mathbf{V}_{\mathbf{A}}\mathbf{T}^H\right\}\nonumber\\
		&=\gamma^2
		\left[ (\mathbf{a}^H\mathbf{V}_{\mathbf{A}}\mathbf{T}^H\mathbf{R}_{\mathbf{y}}^{-1}\mathbf{T}\mathbf{V}_{\mathbf{A}}^H\dot{\mathbf{a}})^2
		+2(\mathbf{a}^H\mathbf{V}_{\mathbf{A}}\mathbf{T}^H\mathbf{R}_{\mathbf{y}}^{-1}\mathbf{T}\mathbf{V}_{\mathbf{A}}^H\mathbf{a}) (\dot{\mathbf{a}}^H\mathbf{V}_{\mathbf{A}}\mathbf{T}^H\mathbf{R}_{\mathbf{y}}^{-1}\mathbf{T}\mathbf{V}_{\mathbf{A}}^H\dot{\mathbf{a}}) +(\dot{\mathbf{a}}^H\mathbf{V}_{\mathbf{A}}\mathbf{T}^H\mathbf{R}_{\mathbf{y}}^{-1}\mathbf{T}\mathbf{V}_{\mathbf{A}}^H\mathbf{a})^2
		\right]\nonumber\\
		&=\gamma^2( F_a+2F_b+F_c)
	\end{align}
\end{figure*}
where
\begin{equation}\label{F1}
F_a=(\mathbf{a}^H\mathbf{V}_{\mathbf{A}}\mathbf{T}^H\mathbf{R}_{\mathbf{y}}^{-1}\mathbf{T}\mathbf{V}_{\mathbf{A}}^H\mathbf{a}')^2,
\end{equation}
\begin{equation}\label{F2}
F_b=(\mathbf{a}^H\mathbf{V}_{\mathbf{A}}\mathbf{T}^H\mathbf{R}_{\mathbf{y}}^{-1}\mathbf{T}\mathbf{V}_{\mathbf{A}}^H\mathbf{a}) (\mathbf{a}'^H\mathbf{V}_{\mathbf{A}}\mathbf{T}^H\mathbf{R}_{\mathbf{y}}^{-1}\mathbf{T}\mathbf{V}_{\mathbf{A}}^H\mathbf{a}')
\end{equation}
and
\begin{equation}\label{F3}
F_c=(\mathbf{a}'^H\mathbf{V}_{\mathbf{A}}\mathbf{T}^H\mathbf{R}_{\mathbf{y}}^{-1}\mathbf{T}\mathbf{V}_{\mathbf{A}}^H\mathbf{a})^2.
\end{equation}
Then, the component of $F_a$ can be simplified as (\ref{Fasub}), where $\mathbf{D}$ can be rewritten in Kronecker product form as
\begin{equation}\label{Dkp}
\mathbf{D}=\mathbf{I}_{M_s}\otimes \mathbf{D}_{a}+\mathbf{D}_{s}\otimes \mathbf{I}_{M_a}
\end{equation}
where
\begin{equation}\label{Da}
\mathbf{D}_{a}=\left[
\begin{array}{cccc}
d_{1,1} & 0         & \cdots & 0 \\
0       & d_{1,2}   & \cdots & 0 \\
\vdots  & \vdots    & \ddots & \vdots \\
0       & 0         & \cdots & d_{1,M_a}
\end{array}
\right]
\end{equation}
and
\begin{equation}\label{Ds}
\mathbf{D}_{s}=\left[
\begin{array}{cccc}
d_{1,1} & 0         & \cdots & 0 \\
0       & d_{2,1}   & \cdots & 0 \\
\vdots  & \vdots    & \ddots & \vdots \\
0       & 0         & \cdots & d_{M_s,1}
\end{array}
\right].
\end{equation}
Thus, $\mathbf{a}^{H}\mathbf{B}\mathbf{D}\mathbf{a}$ can be given by
\begin{align}\label{abda}
\mathbf{a}^{H}&\mathbf{B}\mathbf{D}\mathbf{a}\nonumber\\
&=\frac{1}{M_a}(\mathbf{a}_s\otimes \mathbf{a}_a)^H(\mathbf{C}\otimes\mathbf{1}_{M_a}\mathbf{1}_{M_a}^H)(\mathbf{I}_{M_s}\otimes \mathbf{D}_{a} \nonumber\\
&~~~+\mathbf{D}_{s}\otimes \mathbf{I}_{M_a})(\mathbf{a}_s\otimes \mathbf{a}_a)\nonumber\\
&=\frac{1}{M_a}(\mathbf{a}_s^H\mathbf{C}\mathbf{a}_s\otimes\mathbf{a}_a^H\mathbf{1}_{M_a}\mathbf{1}_{M_a}^H\mathbf{D}_{a}\mathbf{a}_a \nonumber\\ &~~~+\mathbf{a}_s^H\mathbf{C}\mathbf{D}_{s}\mathbf{a}_s\otimes\mathbf{a}_a^H\mathbf{1}_{M_a}\mathbf{1}_{M_a}^H\mathbf{a}_a) \nonumber\\ &=\frac{1}{M_a}\left[\xi\zeta^H\sum_{m_a=1}^{M_a}d_{1,ma}e^{j\frac{2\pi}{\lambda}d_{1,m_a}\sin\theta_0}\right. \nonumber\\
&~~~\left.+\left(\sum_{m_s=1}^{M_0}d_{m_s,1}+\frac{\alpha^2}{\alpha^2+\sigma_q^2}\sum_{m_s=M_0+1}^{M_s}d_{m_s,1}\right)\|\zeta\|^2\right]\nonumber\\
&=\frac{1}{M_a}(\xi\zeta^H\Gamma+\mu\|\zeta\|^2),
\end{align}
where
\begin{equation}\label{Gamma}
\Gamma=\sum_{m_a=1}^{M_a}d_{1,m_a}e^{j\frac{2\pi}{\lambda}d_{1,m_a}\sin\theta_0}
\end{equation}
and
\begin{equation}\label{muh}
\mu=\sum_{m_s=1}^{M_0}d_{m_s,1}+\frac{\alpha^2}{\alpha^2+\sigma_q^2}\sum_{m_s=M_0+1}^{M_s}d_{m_s,1}.
\end{equation}
Substituting (\ref{aba}) and (\ref{abda}) into (\ref{Fasub}) yields
\begin{align}\label{Fdfinal}
F_a&=(\mathbf{a}^H\mathbf{V}_{\mathbf{A}}\mathbf{T}^H\mathbf{R}_{\mathbf{y} }^{-1} \mathbf{T}\mathbf{V}_{\mathbf{A}}^H\mathbf{a}')^2 \nonumber\\
&=-\frac{4\pi^2\cos^2\theta_0(\xi\zeta^H\Gamma+\mu\|\zeta\|^2)^2} {\lambda^2(\gamma\xi\|\zeta\|^2+M_a)^2}.
\end{align}
Furthermore, we can represent the $\mathbf{a}^{H}\mathbf{D}\mathbf{B}\mathbf{a}$ as follows
\begin{align}\label{adba}
\mathbf{a}^{H}&\mathbf{D}\mathbf{B}\mathbf{a}\nonumber\\
&=\frac{1}{M_a}(\mathbf{a}_s\otimes \mathbf{a}_a)^H(\mathbf{I}_{M_s}\otimes \mathbf{D}_{a}+\mathbf{D}_{s}\otimes \mathbf{I}_{M_a}) \nonumber\\
&~~~(\mathbf{C}\otimes\mathbf{1}_{M_a}\mathbf{1}_{M_a}^H)(\mathbf{a}_s\otimes\mathbf{a}_a)\nonumber\\
&=\frac{1}{M_a}(\mathbf{a}_s^H\mathbf{C}\mathbf{a}_s\otimes\mathbf{a}_a^H\mathbf{D}_{a}\mathbf{1}_{M_a}\mathbf{1}_{M_a}^H\mathbf{a}_a +\mathbf{a}_s^H\mathbf{D}_{s}\mathbf{C}\mathbf{a}_s\nonumber\\ &~~~\otimes\mathbf{a}_a^H\mathbf{1}_{M_a}\mathbf{1}_{M_a}^H\mathbf{a}_a) \nonumber\\
&=\frac{1}{M_a}(\xi\zeta\Gamma^H+\mu \|\zeta\|^2).
\end{align}
Hence, $F_c$ is given by
\begin{align}\label{Fffinal}
F_c&=(\mathbf{a}'^H\mathbf{V}_{\mathbf{A}}\mathbf{T}^H\mathbf{R}_{\mathbf{y} }^{-1} \mathbf{T}\mathbf{V}_{\mathbf{A}}^H\mathbf{a})^2 \nonumber\\
&=-\frac{4\pi^2\cos^2\theta_0(\xi\zeta\Gamma^H+\mu\|\zeta\|^2)^2} {\lambda^2(\gamma\xi\|\zeta\|^2+M_a)^2}.
\end{align}
Now, to obtain the $F_b$, we should derive the expression of $\mathbf{a}^H\mathbf{V}_{\mathbf{A}}\mathbf{T}^H\mathbf{R}_{\mathbf{y} }^{-1} \mathbf{T}\mathbf{V}_{\mathbf{A}}^H\mathbf{a}$ firstly, which can be calculated by
\begin{align}\label{Fesub1}
\mathbf{a}^H&\mathbf{V}_{\mathbf{A}}\mathbf{T}^H\mathbf{R}_{\mathbf{y} }^{-1} \mathbf{T}\mathbf{V}_{\mathbf{A}}^H\mathbf{a}\nonumber\\
&=\mathbf{a}^H\mathbf{V}_{\mathbf{A}}\mathbf{T}^H \left(\mathbf{Q} ^{-1} -\frac{\mathbf{Q} ^{-1}\mathbf{T}\mathbf{V}_A ^H\mathbf{a}\mathbf{a}^H\mathbf{V}_A\mathbf{T}^H\mathbf{Q} ^{-1}}{\gamma^{-1}+ \mathbf{a}^{H}\mathbf{B}\mathbf{a}} \right) \nonumber\\
&~~~\times \mathbf{T}\mathbf{V}_{\mathbf{A}}^H\mathbf{a}\nonumber\\
&=\mathbf{a}^{H}\mathbf{B}\mathbf{a}-\frac{(\mathbf{a}^{H}\mathbf{B}\mathbf{a})^2}{\gamma^{-1}+\mathbf{a}^{H}\mathbf{B}\mathbf{a}}\nonumber\\
&=\frac{\xi\|\zeta\|^2}{\gamma\xi\|\zeta\|^2+M_a}
\end{align}
Afterwards, the second item of $F_b$ is derived as
\begin{align}\label{Fesub2}
\mathbf{a}'^H&\mathbf{V}_{\mathbf{A}}\mathbf{T}^H\mathbf{R}_{\mathbf{y} }^{-1} \mathbf{T}\mathbf{V}_{\mathbf{A}}^H\mathbf{a}'\nonumber\\
&=\mathbf{a}'^H\mathbf{V}_{\mathbf{A}}\mathbf{T}^H \left(\mathbf{Q} ^{-1} -\frac{\mathbf{Q} ^{-1}\mathbf{T}\mathbf{V}_A ^H\mathbf{a}\mathbf{a}^H\mathbf{V}_A\mathbf{T}^H\mathbf{Q} ^{-1}}{\gamma^{-1}+ \mathbf{a}^{H}\mathbf{B}\mathbf{a}} \right) \nonumber\\
&~~~\times \mathbf{T}\mathbf{V}_{\mathbf{A}}^H\mathbf{a}'\nonumber\\
&=\frac{4 \pi^2}{\lambda^2}\cos^2\theta_0\left[ \mathbf{a}^{H}\mathbf{D}\mathbf{B}\mathbf{D}\mathbf{a}- \frac{(\mathbf{a}^{H}\mathbf{D}\mathbf{B}\mathbf{a})(\mathbf{a}^{H}\mathbf{B}\mathbf{D}\mathbf{a})}{\gamma^{-1}+\mathbf{a}^{H}\mathbf{B}\mathbf{a}} \right].
\end{align}
Thus, we just need to derive the $\mathbf{a}^{H}\mathbf{D}\mathbf{B}\mathbf{D}\mathbf{a}$, which can be given by
\begin{align}\label{adbda}
\mathbf{a}^{H}&\mathbf{D}\mathbf{B}\mathbf{D}\mathbf{a}\nonumber\\
&=\frac{1}{M_a}(\mathbf{a}_s\otimes \mathbf{a}_a)^H(\mathbf{I}_{M_s}\otimes \mathbf{D}_{a}+\mathbf{D}_{s}\otimes \mathbf{I}_{M_a}) \nonumber\\
&~~~(\mathbf{C}\otimes\mathbf{1}_{M_a}\mathbf{1}_{M_a}^H)(\mathbf{I}_{M_s}\otimes \mathbf{D}_{a}+\mathbf{D}_{s}\otimes \mathbf{I}_{M_s})(\mathbf{a}_s\otimes\mathbf{a}_a)\nonumber\\
&=\frac{1}{M_a}(\mathbf{a}_s^H\mathbf{C}\mathbf{a}_s\otimes\mathbf{a}_a^H\mathbf{D}_{a}\mathbf{1}_{M_a}\mathbf{1}_{M_a}^H\mathbf{D}_{a}\mathbf{a}_a +\mathbf{a}_s^H\mathbf{C}\mathbf{D}_{s}\mathbf{a}_{s} \nonumber\\
&~~~\otimes\mathbf{a}_a^H\mathbf{D}_{a}\mathbf{1}_{M_a}\mathbf{1}_{M_a}^H\mathbf{a}_a +\mathbf{a}_s^H\mathbf{D}_{s}\mathbf{C}\mathbf{a}_{s}\otimes\mathbf{a}_a^H\mathbf{1}_{M_a}\mathbf{1}_{M_a}^H\mathbf{D}_{a}\mathbf{a}_a\nonumber\\
&~~~+\mathbf{a}_s^H\mathbf{D}_{s}\mathbf{C}\mathbf{D}_{s}\mathbf{a}_{s}\otimes\mathbf{a}_a^H\mathbf{1}_{M_a}\mathbf{1}_{M_a}^H\mathbf{a}_a)\nonumber\\
&=\frac{1}{M_a}\left[\xi\Gamma^H\Gamma+\mu(\Gamma^H\zeta +\zeta^H\Gamma)+\left(\sum_{m_s=1}^{M_0}d_{m_s,1}^2\right.\right.\nonumber\\
&~~~\left.\left.+\frac{\alpha^2}{\alpha^2+\sigma_q^2}\sum_{m_s=M_0+1}^{M_s}d_{m_s,1}^2\right)\|\zeta\|^2\right]\nonumber\\
&=\frac{1}{M_a}\left(\xi\|\Gamma\|^2+2\mu\mathbb{R}\left[\Gamma^H\zeta\right]+\nu\|\zeta\|^2\right),
\end{align}
where
\begin{equation}\label{nuh}
\nu=\sum_{m_s=1}^{M_0}d_{m_s,1}^2+\frac{\alpha^2}{\alpha^2+\sigma_q^2}\sum_{m_s=M_0+1}^{M_s}d_{m_s,1}^2.
\end{equation}
Therefore, combining (\ref{Fesub1}), (\ref{Fesub2}) and (\ref{adbda}), we have
\begin{align}\label{Fefinal}
&F_b\nonumber\\
&~~=(\mathbf{a}^H\mathbf{V}_{\mathbf{A}}\mathbf{T}^H\mathbf{R}_{\mathbf{y} }^{-1} \mathbf{T}\mathbf{V}_{\mathbf{A}}^H\mathbf{a}) (\mathbf{a}'^H\mathbf{V}_{\mathbf{A}}\mathbf{T}^H\mathbf{R}_{\mathbf{y} }^{-1} \mathbf{T}\mathbf{V}_{\mathbf{A}}^H\mathbf{a}') \nonumber\\
&~~=\left.\frac{4\pi^2\xi\|\zeta\|^2}{\lambda^2M_a(\gamma\xi\|\zeta\|^2+M_a)}\cos^2\theta_0 \right(\xi\|\Gamma\|^2+2\mu\mathbb{R}\left[\Gamma^H\zeta\right]\nonumber\\
&~~~~\left.+\nu\|\zeta\|^2- \frac{\xi^2\|\Gamma\|^2\|\zeta\|^2 +2\xi\mu\|\zeta\|^2\mathbb{R}\left[\Gamma^H\zeta\right]+\mu^2\|\zeta\|^4}{\gamma^{-1}M_a +\xi\|\zeta\|^2}\right)\nonumber\\
&~~=\left.\frac{4\pi^2\cos^2\theta_0\xi\|\zeta\|^2}{\lambda^2(\gamma\xi\|\zeta\|^2+M_a)^2}\right[ \xi\|\Gamma\|^2+2\mu\mathbb{R}\left[\Gamma^H\zeta\right] \nonumber\\
&~~~~~\left.+\nu\|\zeta\|^2 +\frac{\gamma\|\zeta\|^4}{M_a}\left(\xi\nu-\mu^2\right)\right].
\end{align}

Finally, substitute the (\ref{Fdfinal}), (\ref{Fffinal}) and (\ref{Fefinal}) into (\ref{Fhe}), $\mathbf{F}_{\theta_0,\theta_0}$ is given by
\begin{align}\label{Fthetathetafinal}
\mathbf{F}_{\theta_0,\theta_0}&=\gamma^2(F_a+ 2F_b+F_c) \nonumber\\
&=\frac{8\pi^2\gamma^2\cos^2\theta_0}{\lambda^2M_a(\gamma\xi\|\zeta\|^2+M_a)^2}\nonumber\\
&~~~\times\left[\|\zeta\|^4(\xi\nu-\mu^2)(\gamma\xi\|\zeta\|^2 +M_a)\frac{}{}\right.\nonumber\\
&~~~\left.+M_a\xi^2\left(\|\zeta\|^2\|\Gamma\|^2-\mathbb{R}\left[\left(\Gamma^H\zeta\right)^2\right]\right)\right].
\end{align}
%
%
Now, we complete the derivation for the closed-form expression of the FIM for the HAD architecture with mixed-ADCs.
$\hfill\blacksquare$

\ifCLASSOPTIONcaptionsoff
  \newpage
\fi

\bibliographystyle{IEEEtran}
\bibliography{hybrid_mixedADC_ref}

\begin{thebibliography}{10}
\providecommand{\url}[1]{#1}
\csname url@samestyle\endcsname
\providecommand{\newblock}{\relax}
\providecommand{\bibinfo}[2]{#2}
\providecommand{\BIBentrySTDinterwordspacing}{\spaceskip=0pt\relax}
\providecommand{\BIBentryALTinterwordstretchfactor}{4}
\providecommand{\BIBentryALTinterwordspacing}{\spaceskip=\fontdimen2\font plus
\BIBentryALTinterwordstretchfactor\fontdimen3\font minus
  \fontdimen4\font\relax}
\providecommand{\BIBforeignlanguage}[2]{{%
\expandafter\ifx\csname l@#1\endcsname\relax
\typeout{** WARNING: IEEEtran.bst: No hyphenation pattern has been}%
\typeout{** loaded for the language `#1'. Using the pattern for}%
\typeout{** the default language instead.}%
\else
\language=\csname l@#1\endcsname
\fi
#2}}
\providecommand{\BIBdecl}{\relax}
\BIBdecl

\bibitem{DOA}
T.~E. {Tuncer} and B.~{Friedlander}, \emph{Classical and Modern
  Direction-of-Arrival Estimation}.\hskip 1em plus 0.5em minus 0.4em\relax
  Academic Press, 2009.

\bibitem{chenDOA2022wcl}
Y.~Chen, X.~Zhan, F.~Shu, Q.~Jie, X.~Cheng, Z.~Zhuang, and J.~Wang, ``Two
  low-complexity {DOA} estimators for massive/ultra-massive {MIMO} receive
  array,'' \emph{IEEE Wireless Commun. Lett}, vol.~11, no.~11, pp. 2385--2389,
  2022.

\bibitem{zhuang2020machine}
Z.~Zhuang, L.~Xu, J.~Li, J.~Hu, L.~Sun, F.~Shu, and J.~Wang,
  ``Machine-learning-based high-resolution {DOA} measurement and robust
  directional modulation for hybrid analog-digital massive {MIMO}
  transceiver,'' \emph{Science China Information Sciences}, vol.~63, pp. 1--18,
  2020.

\bibitem{zengUAV}
Y.~{Zeng} and R.~{Zhang}, ``Energy-efficient {UAV} communication with
  trajectory optimization,'' \emph{IEEE Trans. Wireless Commun.}, vol.~16,
  no.~6, pp. 3747--3760, 2017.

\bibitem{shuDMRIS2021tcom}
F.~Shu, Y.~Teng, J.~Li, M.~Huang, W.~Shi, J.~Li, Y.~Wu, and J.~Wang, ``Enhanced
  secrecy rate maximization for directional modulation networks via {IRS},''
  \emph{IEEE Trans. Commun.}, vol.~69, no.~12, pp. 8388--8401, 2021.

\bibitem{HongIRS}
S.~{Hong}, C.~{Pan}, H.~{Ren}, K.~{Wang}, and A.~{Nallanathan},
  ``Artificial-noise-aided secure {MIMO} wireless communications via
  intelligent reflecting surface,'' \emph{IEEE Trans. Commun.}, vol.~68,
  no.~12, pp. 7851--7866, 2020.

\bibitem{WuIRS}
Q.~{Wu} and R.~{Zhang}, ``Beamforming optimization for wireless network aided
  by intelligent reflecting surface with discrete phase shifts,'' \emph{IEEE
  Trans. Commun.}, vol.~68, no.~3, pp. 1838--1851, 2020.

\bibitem{KaurIOT}
N.~{Kaur} and S.~K. {Sood}, ``An energy-efficient architecture for the internet
  of things ({IoT}),'' \emph{IEEE Syst. J.}, vol.~11, no.~2, pp. 796--805,
  2017.

\bibitem{dongRIS2022ojcs}
R.~Dong, S.~Jiang, X.~Hua, Y.~Teng, F.~Shu, and J.~Wang, ``Low-complexity joint
  phase adjustment and receive beamforming for directional modulation networks
  via {IRS},'' \emph{IEEE Open Journal of the Communications Society}, vol.~3,
  pp. 1234--1243, 2022.

\bibitem{jieDOA2022wcl}
Q.~Jie, X.~Zhan, F.~Shu, Y.~Ding, B.~Shi, Y.~Li, and J.~Wang,
  ``High-performance passive eigen-model-based detectors of single emitter
  using massive {MIMO} receivers,'' \emph{IEEE Wireless Commun. Lett}, vol.~11,
  no.~4, pp. 836--840, 2022.

\bibitem{MUSIC}
R.~Schmidt, ``A signal subspace approach to multiple emitter location and
  spectral estimation,'' \emph{Ph. D. Dissertation. Stanford Univ.}, 1981.

\bibitem{rootmusic}
A.~{Barabell}, ``Improving the resolution performance of eigenstructure-based
  direction-finding algorithms,'' in \emph{ICASSP '83. IEEE International
  Conference on Acoustics, Speech, and Signal Processing}, vol.~8, 1983, pp.
  336--339.

\bibitem{ESPRIT}
R.~{Roy}, A.~{Paulraj}, and T.~{Kailath}, ``{ESPRIT}--a subspace rotation
  approach to estimation of parameters of cisoids in noise,'' \emph{IEEE
  Transactions on Acoustics, Speech, and Signal Processing}, vol.~34, no.~5,
  pp. 1340--1342, 1986.

\bibitem{TLS_ESPRIT}
R.~Roy and T.~Kailath, ``Esprit-estimation of signal parameters via rotational
  invariance techniques,'' \emph{IEEE Transactions on Acoustics, Speech, and
  Signal Processing}, vol.~37, no.~7, pp. 984--995, 1989.

\bibitem{U_rootmusic}
M.~{Pesavento}, A.~B. {Gershman}, and M.~{Haardt}, ``Unitary root-{MUSIC} with
  a real-valued eigendecomposition: a theoretical and experimental performance
  study,'' \emph{IEEE Trans. Signal Process.}, vol.~48, no.~5, pp. 1306--1314,
  2000.

\bibitem{RV_rootmusic}
F.~{Yan}, M.~{Jin}, S.~{Liu}, and X.~{Qiao}, ``Real-valued {MUSIC} for
  efficient direction estimation with arbitrary array geometries,'' \emph{IEEE
  Trans. Signal Process.}, vol.~62, no.~6, pp. 1548--1560, 2014.

\bibitem{chengDOA}
L.~{Cheng}, Y.~{Wu}, J.~{Zhang}, and L.~{Liu}, ``Subspace identification for
  {DOA} estimation in massive/full-dimension {MIMO} systems: Bad data
  mitigation and automatic source enumeration,'' \emph{IEEE Trans. Signal
  Process.}, vol.~63, no.~22, pp. 5897--5909, 2015.

\bibitem{LeeADC}
H.~{Lee} and C.~G. {Sodini}, ``Analog-to-digital converters: Digitizing the
  analog world,'' \emph{Proc. IEEE}, vol.~96, no.~2, pp. 323--334, 2008.

\bibitem{shiHADDOA2022scis}
B.~Shi, X.~Jiang, N.~Chen, Y.~Teng, J.~Lu, F.~Shu, J.~Zou, J.~Li, and J.~Wang,
  ``Fast ambiguous {DOA} elimination method of {DOA} measurement for hybrid
  massive {MIMO} receiver,'' \emph{SCIENCE CHINA Information Sciences},
  vol.~65, no.~5, pp. 159\,302--, 2022.

\bibitem{shuDOA}
F.~{Shu}, Y.~{Qin}, T.~{Liu}, L.~{Gui}, Y.~{Zhang}, J.~{Li}, and Z.~{Han},
  ``Low-complexity and high-resolution {DOA} estimation for hybrid analog and
  digital massive {MIMO} receive array,'' \emph{IEEE Trans. Commun.}, vol.~66,
  no.~6, pp. 2487--2501, 2018.

\bibitem{wuDOA}
K.~{Wu}, W.~{Ni}, T.~{Su}, R.~P. {Liu}, and Y.~J. {Guo}, ``Robust unambiguous
  estimation of angle-of-arrival in hybrid array with localized analog
  subarrays,'' \emph{IEEE Trans. Wireless Commun.}, vol.~17, no.~5, pp.
  2987--3002, 2018.

\bibitem{fanDOA}
D.~{Fan}, F.~{Gao}, Y.~{Liu}, Y.~{Deng}, G.~{Wang}, Z.~{Zhong}, and
  A.~{Nallanathan}, ``Angle domain channel estimation in hybrid millimeter wave
  massive {MIMO} systems,'' \emph{IEEE Trans. Wireless Commun.}, vol.~17,
  no.~12, pp. 8165--8179, 2018.

\bibitem{LiDOA}
S.~Li, Y.~Liu, L.~You, W.~Wang, H.~Duan, and X.~Li, ``Covariance matrix
  reconstruction for doa estimation in hybrid massive mimo systems,''
  \emph{IEEE Wireless Commun. Lett}, vol.~9, no.~8, pp. 1196--1200, 2020.

\bibitem{huDLDOA}
D.~{Hu}, Y.~{Zhang}, L.~{He}, and J.~{Wu}, ``Low-complexity deep-learning-based
  {DOA} estimation for hybrid massive {MIMO} systems with uniform circular
  arrays,'' \emph{IEEE Wireless Commun. Lett.}, vol.~9, no.~1, pp. 83--86,
  2020.

\bibitem{SinghLowADC}
J.~{Singh}, O.~{Dabeer}, and U.~{Madhow}, ``On the limits of communication with
  low-precision analog-to-digital conversion at the receiver,'' \emph{IEEE
  Trans. Commun.}, vol.~57, no.~12, pp. 3629--3639, 2009.

\bibitem{FanLowADC}
L.~{Fan}, S.~{Jin}, C.~{Wen}, and H.~{Zhang}, ``Uplink achievable rate for
  massive {MIMO} systems with low-resolution {ADC},'' \emph{IEEE Commun.
  Lett.}, vol.~19, no.~12, pp. 2186--2189, 2015.

\bibitem{JinLowADC}
S.~{Jin}, X.~{Liang}, K.~{Wong}, X.~{Gao}, and Q.~{Zhu}, ``Ergodic rate
  analysis for multipair massive {MIMO} two-way relay networks,'' \emph{IEEE
  Trans. Wireless Commun.}, vol.~14, no.~3, pp. 1480--1491, 2015.

\bibitem{KongLowADC}
C.~{Kong}, C.~{Zhong}, S.~{Jin}, S.~{Yang}, H.~{Lin}, and Z.~{Zhang},
  ``Full-duplex massive {MIMO} relaying systems with low-resolution {ADCs},''
  \emph{IEEE Trans. Wireless Commun.}, vol.~16, no.~8, pp. 5033--5047, 2017.

\bibitem{XuLOWADC}
L.~{Xu}, L.~{Sun}, G.~{Xia}, T.~{Liu}, F.~{Shu}, Y.~{Zhang}, and J.~{Wang},
  ``Secure hybrid digital and analog precoder for {mmWave} systems with
  low-resolution {DACs} and finite-quantized phase shifters,'' \emph{IEEE
  Access}, vol.~7, pp. 109\,763--109\,775, 2019.

\bibitem{wangDOA2018twc}
C.-J. Wang, C.-K. Wen, S.~Jin, and S.-H. Tsai, ``Gridless channel estimation
  for mixed one-bit antenna array systems,'' \emph{IEEE Trans. Wireless
  Commun.}, vol.~17, no.~12, pp. 8485--8501, 2018.

\bibitem{Shiarxiv}
B.~Shi, N.~Chen, X.~Zhu, Y.~Qian, Y.~Zhang, F.~Shu, and J.~Wang, ``Impact of
  low-resolution {ADC} on {DOA} estimation performance for massive {MIMO}
  receive array,'' \emph{IEEE Syst. J.}, vol.~16, no.~2, pp. 2635--2638, 2022.

\bibitem{LiangMIXED}
N.~{Liang} and W.~{Zhang}, ``Mixed-{ADC} massive {MIMO},'' \emph{IEEE J. Sel.
  Areas Commun.}, vol.~34, no.~4, pp. 983--997, 2016.

\bibitem{Zhangtcom}
J.~{Zhang}, L.~{Dai}, Z.~{He}, B.~{Ai}, and O.~A. {Dobre}, ``Mixed-{ADC/DAC}
  multipair massive {MIMO} relaying systems: Performance analysis and power
  optimization,'' \emph{IEEE Trans. Commun.}, vol.~67, no.~1, pp. 140--153,
  2019.

\bibitem{accessMIXED}
W.~{Tan}, S.~{Jin}, C.~{Wen}, and Y.~{Jing}, ``Spectral efficiency of
  mixed-{ADC} receivers for massive {MIMO} systems,'' \emph{IEEE Access},
  vol.~4, pp. 7841--7846, 2016.

\bibitem{shiDOAmixedADC}
B.~Shi, L.~Zhu, W.~Cai, N.~Chen, T.~Shen, P.~Zhu, F.~Shu, and J.~Wang, ``On
  performance loss of {DOA} measurement using massive {MIMO} receiver with
  mixed-{ADC}s,'' \emph{IEEE Wireless Commun. Lett}, vol.~11, no.~8, pp.
  1614--1618, 2022.

\bibitem{Lowpower}
O.~{Orhan}, E.~{Erkip}, and S.~{Rangan}, ``Low power analog-to-digital
  conversion in millimeter wave systems: Impact of resolution and bandwidth on
  performance,'' in \emph{2015 Information Theory and Applications Workshop
  (ITA)}, 2015, pp. 191--198.

\bibitem{Zhangjsac}
J.~{Zhang}, L.~{Dai}, Z.~{He}, S.~{Jin}, and X.~{Li}, ``Performance analysis of
  mixed-{ADC} massive {MIMO} systems over rician fading channels,'' \emph{IEEE
  J. Sel. Areas Commun.}, vol.~35, no.~6, pp. 1327--1338, 2017.

\bibitem{jinshilow}
J.~{Zhu}, C.~{Wen}, J.~{Tong}, C.~{Xu}, and S.~{Jin}, ``Grid-less variational
  bayesian channel estimation for antenna array systems with low resolution
  {ADC}s,'' \emph{IEEE Trans. Wireless Commun.}, vol.~19, no.~3, pp.
  1549--1562, 2020.

\bibitem{CuiEnergy}
S.~{Cui}, A.~J. {Goldsmith}, and A.~{Bahai}, ``Energy-constrained modulation
  optimization,'' \emph{IEEE Trans. Wireless Commun.}, vol.~4, no.~5, pp.
  2349--2360, 2005.

\bibitem{GaoHAD}
X.~{Gao}, L.~{Dai}, S.~{Han}, C.~{I}, and R.~W. {Heath}, ``Energy-efficient
  hybrid analog and digital precoding for mmwave {MIMO} systems with large
  antenna arrays,'' \emph{IEEE J. Sel. Areas Commun.}, vol.~34, no.~4, pp.
  998--1009, 2016.

\bibitem{HeEnergy}
C.~{He}, B.~{Sheng}, P.~{Zhu}, and X.~{You}, ``Energy efficient comparison
  between distributed {MIMO} and co-located {MIMO} in the uplink cellular
  systems,'' in \emph{2012 IEEE Vehicular Technology Conference (VTC Fall)},
  2012, pp. 1--5.

\bibitem{Matrix}
R.~A. {Horn} and C.~R. {Johnson}, \emph{Matrix Analysis}.\hskip 1em plus 0.5em
  minus 0.4em\relax Cambridge University Press, 2012.

\end{thebibliography}

\end{document}